\documentclass[Afour,sageh,times]{sagej}
\usepackage[utf8]{inputenc}
\usepackage{graphicx}        % standard LaTeX graphics tool
\usepackage{courier}
\usepackage{listings}
\usepackage{color}
\usepackage{hyperref}
\usepackage{breakurl}
\usepackage{array}
\usepackage{amssymb}
\usepackage{multicol}
\newcommand{\GHcomm}[1]{{\color{black}#1\color{black}}}
\usepackage{algorithm}
\usepackage[noend]{algpseudocode}
\newcommand{\hf}{\textcolor{black}}
\newcommand{\gh}{\textcolor{black}}
\newcommand{\final}{\textcolor{black}}

\makeatletter
\def\BState{\State\hskip-\ALG@thistlm}
\makeatother

\usepackage{amsmath}
\usepackage{amsfonts}
\usepackage{amssymb}
\usepackage{bbm}
\usepackage{braket}

\newcommand{\bq}{\begin{equation}}
\newcommand{\eq}{\end{equation}}
\newcommand{\second}{\mbox{sec}}
\newcommand{\flop}{\mbox{flop}}
\newcommand{\flops}{\mbox{flops}}

\newcommand{\GBS}{\mbox{G\byte/\second}}

\newcommand{\lup}{\mbox{LUP}}

\newcommand{\GLUPS}{\mbox{G\lup{}s/\second}}
\newcommand{\GRNS}{\mbox{GRN/\second}}
\newcommand{\GHZ}{\mbox{GHz}}

\newcommand{\BF}{\mbox{\bytes/\flop}}
\newcommand{\BL}{\mbox{\bytes/\lup}}
\newcommand{\bytes}{\mbox{bytes}}
\newcommand{\byte}{\mbox{byte}}
\newcommand{\GB}{\mbox{GB}}

\newcommand{\eos}{~.}
\newcommand{\cma}{~,}

\newcommand{\pvsctm}{PVSC-DTM}

\newcommand{\ii}{\mathrm{i}}
\newcommand{\ee}{\mathrm{e}}

\definecolor{mygreen}{rgb}{0,0.6,0}
\definecolor{orange}{rgb}{1.0,0.5,0.0}
\definecolor{purple}{rgb}{0.5,0,0.5}

\lstdefinestyle{customc}{
  belowcaptionskip=1\baselineskip,
  breaklines=true,
  frame=lines,
  xleftmargin=\parindent,
  language=C,
  showstringspaces=false,
  basicstyle=\footnotesize\ttfamily,
  keywordstyle=\bfseries\color{mygreen},
  commentstyle=\itshape\color{purple},
  identifierstyle=\color{blue},
  stringstyle=\color{orange},
}
\begin{document}
\runninghead{Pieper, Hager, and Fehske}
\title{A domain-specific language and matrix-free
  stencil code for investigating electronic properties of 
  Dirac and topological materials}
%\titlerunning{Stencil-Codes for topological materials}
\author{Andreas Pieper\affilnum{1,2}, Georg Hager\affilnum{2}, and Holger Fehske\affilnum{1}}
\affiliation{\affilnum{1}Universt\"at Greifswald, Germany\\
\affilnum{2}Erlangen Regional Computing Center (RRZE), Friedrich-Alexander-Universit\"at Erlangen-N\"urnberg, Germany}
\corrauth{Georg Hager, Erlangen Regional Computing Center
  Martensstr. 1
  91058 Erlangen
  Germany.}

\email{georg.hager@fau.de}

%\affiliation{Institut f\"ur Physik, Ernst-Moritz-Arndt Universt\"at Greifswald, Felix-Hausdorff-Str. 6, 17489 Greifswald, Germany}
%\author{Georg Hager}
%\email{georg.hager@fau.de}
%\affiliation{Erlangen Regional Computing Center (RRZE), Martensstr. 1, 91058 Erlangen, Germany}
%\author{Holger Fehske}
%\email{fehske@physik.uni-greifswald.de}
%\affiliation{Institut f\"ur Physik, Ernst-Moritz-Arndt Universt\"at Greifswald, Felix-Hausdorff-Str. 6, 17489 Greifswald, Germany}
%\authorrunning{A. Pieper et al.}
%\institute{
%A. Pieper $\cdot$ G. Hager
%\at Regionales Rechenzentrum Erlangen, Friedrich-Alexander Universit\"at Erlangen-N\"urnberg, 
%Martensstr. 1, D-91058 Erlangen, Germany\\
%\email\texttt{\{gerald.schubert,georg.hager\}@rrze.uni-erlangen.de}
%\and
%H. Fehske
%\at
%Ernst-Moritz-Arndt-Universit\"at Greifswald,
%Institut f\"ur Physik, Felix-Hausdorff-Str. 6, \\
%D-17489 Greifswald, Germany\\
%\email \texttt{fehske@physik.uni-greifswald.de}
%}

%failed with: \usepackage{hyperref}
%\maketitle  
\begin{abstract}
  We introduce \pvsctm\ \GHcomm{(Parallel Vectorized Stencil Code for Dirac
  and Topological Materials), a library
  and code generator based on a domain-specific language tailored to
  implement the specific stencil-like algorithms that can describe
  Dirac and topological materials such as graphene and topological
  insulators in a matrix-free way. The generated hybrid-parallel
  (MPI+OpenMP) code is fully vectorized using Single Instruction Multiple
  Data (SIMD) extensions.} It is significantly faster than matrix-based
  approaches on the node level and performs in accordance with the
  roof{}line model. We demonstrate the chip-level performance and
  distributed-memory scalability of basic building blocks such as
  sparse matrix-(multiple-) vector multiplication on modern multicore
  CPUs. As an application example, we use the \pvsctm\ scheme to (i)
  explore the scattering of a Dirac wave on an array of gate-defined
  quantum dots, to (ii) calculate a bunch of interior eigenvalues for
  strong topological insulators, and to (iii) discuss the
  photoemission spectra of a disordered Weyl semimetal.
\end{abstract}
\maketitle

%broadep2
%CPU name:	Intel(R) Xeon(R) CPU E5-2697 v4 @ 2.30GHz
%CPU type:	Intel Xeon Broadwell EN/EP/EX processor
%AVX2
%Sockets:		2
%Cores per socket:	18
%Threads per core:	2
%Level:			1 Size:			32 kB
%Level:			2 Size:			256 kB
%Level:			3 Size:			45 MB
%B = 76,8 GB/s (eff. 60 GB/s)

%\tableofcontents

%\emergencystretch5pt

\section{Introduction and related work}

Dirac-type semimetals and topological insulators are new materials with an enormous application potential in fields ranging 
from nano-electronics, plasmonics and optics to quantum information and computation. Their striking electronic, spectroscopic, and transport properties 
result from spin-polarized (chiral), (semi)metallic surface states, which are located in the middle of the spectrum and show linear dispersion to a good approximation.   
The discovery of such massless Dirac fermions in graphene by~\cite{Graphene_review_geim}, on the surface of topological insulators by~\cite{topi_review_hasan_RevModPhys.82.3045}, and in Weyl semimetals by~\cite{Xu613}   has triggered the investigation of \hf{Dirac physics. Realizing  that  certain transport, magnetic and optical properties of solid state systems can be expressed by topological invariants that are insensitive to local perturbations, has largely changed the focus and direction of current condensed matter research from strong correlation to topological aspects (see~\cite{CGMC17}).}

\hf{Whether a material develops} distinct topological phases is dictated by the dimension, the lattice structure and associated electronic band structure
including the boundary states, and the relevant interactions, all reflected in the system's Hamilton operator and its symmetries. Therefore it is of great interest to determine
and analyze the ground-state and spectral properties of paradigmatic model Hamiltonians for topological matter. This can be achieved by means of unbiased numerical approaches.   

\pvsctm\ is a highly parallel, vectorized \GHcomm{(matrix-free)}  stencil code for investigating the properties of two-dimensional (2D) graphene and graphene-nanoribbons (GNRs),
three-dimensional (3D) topological insulators  as well as  Weyl semimetals, including also disorder effects,  by using modern numerical methods based
on matrix polynomials. Due to the complexity of the problem, a considerable amount of computation is required. 
Thus, one of the design goals of \pvsctm\ was to build highly parallel software that supports the
architectural features of modern computer systems, notably SIMD
(Single Instruction Multiple Data) parallelism, shared-memory thread
parallelism, and massively parallel, distributed-memory
parallelism. On the compute node level, the development process was
guided by performance models to ensure that the relevant bottleneck is
saturated. The major methodological advantage compared to existing
software packages for similar purposes is that all matrix operations
are performed without an explicitly stored matrix, thereby greatly 
reducing the pressure on the memory interface and opening 
possibilities for advanced optimizations developed for 
stencil-type algorithms.

In order to ease the burden on users and still provide the flexibility
to adapt the code to different physical setups, a domain-specific
language (DSL) was developed that allows for a formulation
of the problem without any reference to a specific implementation,
let alone optimization. The actual code is generated automatically,
including parallelization and blocking optimizations. Although
several stencil DSLs have been developed (see, e.g., \cite{Pochoir,Snowflake,Exaslang}),
some
even with specific application fields in mind such as in \cite{Halide}, there
is to date no domain-specific approach to generating efficient
stencil code for algorithms describing the specific quantum systems
mentioned above from a high-level representation.
\gh{Optimal blocking factors and other optimization strategies
  are traditionally determined using
  auto-tuning, which was extensively analyzed in the past by, e.g.,
  \cite{Datta:2008,Datta:2009,Kamil:2010,Basu:2013}. Here
we calculate optimal blocking
factors automatically from machine properties,
which makes performance tuning (automatically or manually)
on the generated code or within the code generation phase
obsolete. 
The touchstone for performance optimality is whether the
spMVM loop can achieve minimal code balance and still
utilize a large fraction of the memory bandwidth.
In all application cases investigated so far,
this was not observed as a restriction or disadvantage}.

\gh{Temporal blocking strategies have been a subject of intense
  research over the last two decades
  (\cite{Wonnacott:2000,Bandishti6468470,Girih2017}). They perform
  multiple stencil sweeps of in-cache tiles in order to (ideally)
  decouple from the main memory bottleneck. In unmodified form,
  these approaches
  are unsuitable for the applications covered here because spMVM
  is only a part (albeit an important one) of the whole algorithm.
  However, blocking optimizations do exist for, e.g., filter
  diagonalization methods, and can have a similar effect
  as temporal blocking for pure stencil codes, as shown by
  \cite{Kreutzer:2015,Kreutzer:2018}}.

This report gives an overview of the physical motivation and
describes in detail the implementation of the framework, 
including the DSL. 
Performance models are developed to confirm the optimal
resource utilization on the chip level and assess the potential
of code optimizations, such as spatial blocking and on-the-fly
random number generation.
Performance comparisons on the node and the highly parallel
level with matrix-bound techniques (using the GHOST library)
show the benefit of a matrix-free formulation. 
The code is freely available for download at
\url{http://tiny.cc/PVSC-DTM}\@.

For the benchmark tests we used two different com\-pute nodes: A dual-socket
Intel Xeon E5-2660v2 ``Ivy Bridge'' (IVB) node with 10 cores per socket and 2.2\,\GHZ\
of nominal clock speed, and an Intel Xeon  E5-2697v4 ``Broadwell'' (BDW)
node with 18 cores per socket and 2.3\,\GHZ\ of nominal clock speed.
In all cases the clock frequency was fixed to the nominal value (i.e.,
Turbo Boost was not used)\@. The ``cluster on die'' (CoD) mode was switched
off on BDW, so both systems ran with two ccNUMA domains. The maximum
achievable per-socket memory bandwidth was 40\,\GBS\ on IVB and
61\,\GBS\ on BDW\@. The Intel C/C++ compiler in version 16.0 was used
for compiling the source code.

For all distributed-memory benchmarks we employed the ``Emmy'' cluster
at RRZE (Erlangen Regional Computing Center). This cluster comprises over 500
of the IVB nodes described above, each equipped with 64\,\GB\ of RAM
and connected via a full nonblocking
fat-tree InfiniBand network.

This paper is organized as follows. Section~2  provides typical lattice-model Hamiltonians for graphene nanoribbons with imprinted quantum dots,
strong topological insulators, and disordered Weyl semimetals. A matrix-free method and code for the calculation of electronic properties of these topological systems
is described in Sec.~3, with a focus on a domain-specific language that serves as an input to a code generator. To validate and benchmark the performance of the  numerical approach, the proposed \pvsctm\ scheme  
is executed for several test cases.  Section~4 describes the matrix-polynomial algorithms used for some physical ``real-world'' applications. Finally, we conclude in Sec.~5.

\section{Model Hamiltonians}

In this section we specify the microscopic model Hamiltonians under consideration, in a form best suitable for the application of the \pvsctm\ stencil code. The emergence of Dirac-cone physics is demonstrated.

\subsection{Graphene}

Graphene consists of carbon atoms arranged in a 2D honeycomb lattice structure (see the review by \cite{Graphene_review_geim}). The honeycomb lattice is not a Bravais lattice, \hf{because two neighboring sites are inequivalent from a crystallographic point of view, but can be viewed as a triangular lattice with a two-atom basis, as shown in~\cite{Go11}.} 

\hf{Taking into account only nearest-neighbor hopping processes on the honeycomb lattice, the  resulting \emph{two-band} structure  of pure graphene, 
\begin{eqnarray}
  {\varepsilon}_\pm({\bf k}) & = & \pm\Big[ 3+2\cos(\sqrt{3} k_y a) \nonumber\\
    & & {} +4\cos(\tfrac{\sqrt{3}}{2} k_y a)\cos(\tfrac{3}{2}k_xa)\Big]^{1/2}
  \label{graphene_e_k} 
\end{eqnarray}
exhibits an upper (+) anti-bonding $\pi^*$ band and a lower (-) bonding $\pi$ band, which touch each other at so-called Dirac points; next to any of those the dispersion becomes linear (see \cite{Graphene_review_geim}).  In the following, we set the lattice constant $a=1$. The corresponding graphene tight-binding Hamiltonian respects time-inversion symmetry, which implies $\varepsilon(-{\bf k})=\varepsilon({\bf k})$, and if ${\bf k}^D$ is the solution for $\varepsilon({\bf k})=0$ [which is the Fermi energy $E_F$ for intrinsic (undoped) graphene],  so is $-{\bf k}^D$, i.e., the Dirac points occur in pairs.}

Compared to the band structure of an infinite 2D graphene sheet, the DOS of finite %(graphene nanoribbons)
GNRs is characterized by a multitude of Van Hove singularities, as shown by \cite{Graphene_review_geim} and \cite{SSF09}.  For zigzag GNRs, the strong signature at $E=0$ indicates the high degeneracy of edge states, as shown in  Fig.~\ref{fig:dos_TM_clean}~(a). By contrast, armchair GNRs are gapped around $E=0$; this finite-size gap vanishes when the width of the ribbon tends to infinity.

\hf{Particularly with regard to  the implementation of the PVSC-DTM,  the effective tight-binding Hamiltonian }
for graphene's $\pi$-electrons is brought into the form:
\begin{eqnarray}
  H  & =  &   \sum_{n=1}^{N/4} \left( \Psi_{n+\hat {\bf e}_x}^\dagger T_x  {\Psi_{n}}
                                + \Psi_{n+\hat{\bf e}_y}^\dagger T_y  {\Psi_{n}} + \text{H.c.} 
                           \right)\nonumber \\
    & &  {}  + \sum_{n=1}^{N/4}        \Psi_{n           }^\dagger (T_n+V_n)  {\Psi_{n}}\,,
       \label{graphene_model} 
\end{eqnarray}
\hf{where $\Psi_n$ is a four-component spinor at site $n$. Here and in what follows we use units such that $\hbar =1$ and measure the energy in terms of the carbon-carbon electron transfer integral $t$; $N$ is the number of lattice sites. Then, 
in Eq.~(\ref{graphene_model}), the first term describes the particle transfer $T_{x,y}$ between  neighboring cells (containing now \emph{four} atoms each) in $x$ and $y$ direction,  while the second term gives the transfer $T_n$ within the cells.  To include the case of (on-site) disorder, we are allowing the potentials $v_{n,j}$ to vary  within the cells and from cell to cell. Then  4$\times$4 matrices are}     
\begin{align}
  &T_x = 
   \begin{pmatrix}
     0 & 0 & 0 & 0 \\
    -1 & 0 & 0 & 0 \\
     0 & 0 & 0 &-1 \\
     0 & 0 & 0 & 0
   \end{pmatrix}\ , \;
  T_y = 
   \begin{pmatrix}
     0 & 0 & 0 &-1 \\
     0 & 0 & 0 & 0 \\
     0 & 0 & 0 & 0 \\
     0 & 0 & 0 & 0
   \end{pmatrix}\ ,\nonumber \hspace*{1cm}
\\
  &T_n = 
   \begin{pmatrix}
    0  &-1       &    0     & 0               \\
   -1        & 0 &-1       &0                \\
             0&   -1  &  0&-1              \\
             0& 0        &-1       & 0
   \end{pmatrix}\ ,\;\vspace*{2mm}
\\
   & V_n = \begin{pmatrix}
    v_{n,0}  &0       &0         &0                \\
    0        & v_{n,1}&0       &              0  \\
        0     &0      &  v_{n,2}&0             \\
        0     &    0     &0       &v_{n,3} 
   \end{pmatrix}\,.\nonumber
\end{align}

\subsection{Topological insulators}

The remarkable properties of 3D topological insulators (TIs) result from a particular topology of their band structure, which exhibits gapped (i.e., insulating) bulk and gapless (i.e., metallic) linearly dispersed Dirac surface states (see reviews by \cite{topi_review_hasan_RevModPhys.82.3045} and \cite{topi_review_Qi_RevModPhys.83.1057}).  Bulk-surface correspondence implies, as shown by \cite{FKM07}, that so-called weak TIs (which are less robust against the influence of non-magnetic impurities) feature none or an even number of helical Dirac cones while strong (largely robust) ${\mathbb Z}_2$ TIs have  a single Dirac cone.

As a minimal theoretical model for a 3D TI with cubic lattice structure we consider -- inspired by the orbitals of strained 3D HgTe or the insulators of the $\rm Bi_2Se_3$ family, as studied in~\cite{SRAF12} -- the following \hf{\emph{four-band}} Hamiltonian:
\begin{eqnarray}
       H   & = &  \sum_{n=1}^{N} \Psi_{n}^\dagger \left( 
                      m  \,     \Gamma^1 
                    + \Delta_1 \Gamma^5
                    + \Delta_2 \Gamma^{15}
                    + V_n \Gamma^0
                   \right) {\Psi_{n}}\nonumber\\
          &  &  {}-\sum_{n=1}^{N} \sum_{j=1}^{3} \left(  \Psi_{n+\hat {\bf e}_j}^\dagger
                  \frac{\Gamma^1 - \ii \Gamma^{j+1} }{2}\,
                 %\ee^{\ii \tau_{n,j}} \,  & hoppind disorder
                \Psi_{n} + \text{H.c.}\right), 
               \label{eq:TB_H_topi}
\end{eqnarray}
where $\Psi_n$ is a four-component spinor at site $n$. %In \eqref{eq:TB_H_topi},
The Hamiltonian is expressed in terms of the five Dirac matrices
$\Gamma^a$, $\Gamma^{(1,2,3,4,5)} = ({1} \otimes s_z, -\sigma_y \otimes s_x, \sigma_x
\otimes s_x, -{1} \otimes s_y, \sigma_z\otimes s_x$), and their ten commutators $\Gamma^{ab}=[\Gamma^a,\Gamma^b]/2i$, which satisfy the Clifford  algebra, $\{\Gamma^a,\Gamma^b\}= 2\delta_{a,b} \Gamma^0$,  with $\Gamma^0$ being the identity $\mathbbm{1}_4$ and $s_i$ ($\sigma_i$) the Pauli matrices referring to orbital (spin) space (see~\cite{SFFV12} and  \cite{pf2016_Tpoi_PhysRevB.93.035123}). Hence, $H$  constitutes a complex, sparse, banded 
matrix with seven subdiagonals of small dense blocks of size $4\times 4$. 
\hf{The corresponding tight-binding $4\times 4$ band matrix reads:}
\begin{eqnarray}
         \varepsilon({\bf k}) & = & - \sum_{j=1}^{3} 
         \left( \Gamma^1 - \ii \Gamma^{j+1} \right) \cos k_j  \nonumber\\
         & & {} + m  \,     \Gamma^1 
                    + \Delta_1 \Gamma^5
                    + \Delta_2 \Gamma^{15}\,.
              %+ H_{\text Z}~.  
              \label{eq:bloch_H_topi}
\end{eqnarray}
The parameter $m$ can be used to tune the band structure: For $|m|<1$, a weak TI with two Dirac cones per surface arises, whereas for $1<|m|<3$, a strong TI  results, with a single Dirac cone per surface (see Fig.~\ref{fig:dos_TM_clean}~(b)). In the case that $|m|>3$ we have a conventional band insulator. 
External magnetic fields cause finite $\Delta_1$ and $\Delta_2$, which will break the inversion symmetry. $\Delta_1$, in addition, breaks the time-inversion symmetry.

We now describe for the TI problem how the density of states (DOS) and the single-particle spectral function $A({\bf k}, E)$ depicted in Fig.~\ref{fig:dos_TM_clean} is obtained using state-of-the art exact diagonalization and kernel polynomial methods that were described in~\cite{WF08a} and~\cite{kpm}, respectively. For a given sample geometry (and disorder realization), we can calculate $\{ | l \rangle\}$, the two-fold Kramers degenerate eigenstates 
of $H$~(\ref{eq:bloch_H_topi}).
\hf{Those can be visualized in momentum and  energy space, via the momentum-resolved spectral function 
\begin{equation}
  A({\bf k}, E) = \sum_{\nu=1}^4 \sum_{l=1}^{4N} | \langle l | \psi({\bf k},\nu) \rangle |^2 \delta(E - E_l)
  \label{eq:ak}
\end{equation}
and the density of states (DOS)
\begin{align}
  {\rm DOS}(E) &= \sum_{l=1}^{4N} \delta(E - E_l)\nonumber\\&=  \frac{1}{N}\sum_n\sum\limits_{\nu=1}^{4}\sum\limits_{l=1}^{4N} 
  |\langle l | \psi({\bf r}_n,\nu)\rangle|^2 \delta(E-E_l)\,,
\label{DOS}
\end{align}
even in the case of disorder. 
Here, $\langle {\bf e}^{\,(p)}_n\!\!\otimes\!{\bf e}^{\,(b)}_\nu  | 
\psi({\bf k},\nu')\rangle = 
\exp(\mathrm{i} \langle {\bf  k}|  {\bf e}^{\,(p)}_n  \rangle) \delta_{\nu\nu'}$
is a Bloch state and 
 $\langle|  {\bf e}^{\,(p)}_n\!\!\otimes\!{\bf e}^{\,(b)}_\nu | 
 \psi({\bf r}_n',\nu')\rangle = \delta_{nn'} \delta_{\nu\nu'}$ is a Wannier state, where 
$|{\bf e}^{\,(p)}_n\rangle$ and $|{\bf e}^{\,(b)}_\nu\rangle$
denote the canonical basis vectors of position and band index space, respectively, see~\cite{SFFV12}.}

For the model~\eqref{eq:TB_H_topi} with $m=2$ (and  $V_n=0$, $\Delta_{1/2} =0$), bulk states occur for energies $|E|\geq 1$. Moreover, subgap surface states develop, forming a  
Dirac cone located at the surface momentum ${\bf k}^D=(0,0)$, as shown in Fig.~\ref{fig:dos_TM_clean}~(b). The latter states determine the striking electronic properties of TIs.

\subsection{Weyl semimetals}

The Weyl semimetallic phase, which can be observed, e.g., in TaAS (see \cite{Xu613}), is characterized by a set of linear-dispersive band touching points of two adjacent bands, the so-called Weyl nodes. The real-space Weyl points are associated with chiral fermions, which behave in momentum space like magnetic monopoles. Unlike the 2D Dirac points in graphene, 
the 3D Weyl nodes are protected by the symmetry of the band structure and, as long as there is no translational-symmetry-breaking  intervalley-mixing between different Weyl nodes, the Weyl semimetal is robust against perturbations, as was shown by~\cite{Yang_weyl}. In this way a Weyl semimetal hosts, like a TI, metallic topological surface states (arising from bulk topological invariants). However, while the topological surface states of TIs give rise to a closed Fermi surface (in momentum space), the surface-state band structure of Weyl semimetals is more exotic; it forms open curves, the so-called Fermi arcs, which terminate on the bulk Weyl points (see \cite{Wan_weyl}).  

\begin{figure}%[htbp]
  %\centering\includegraphics[width=0.7\linewidth,clip]{dos_TM_clean}
  \centering\includegraphics[width=1.0\linewidth,clip]{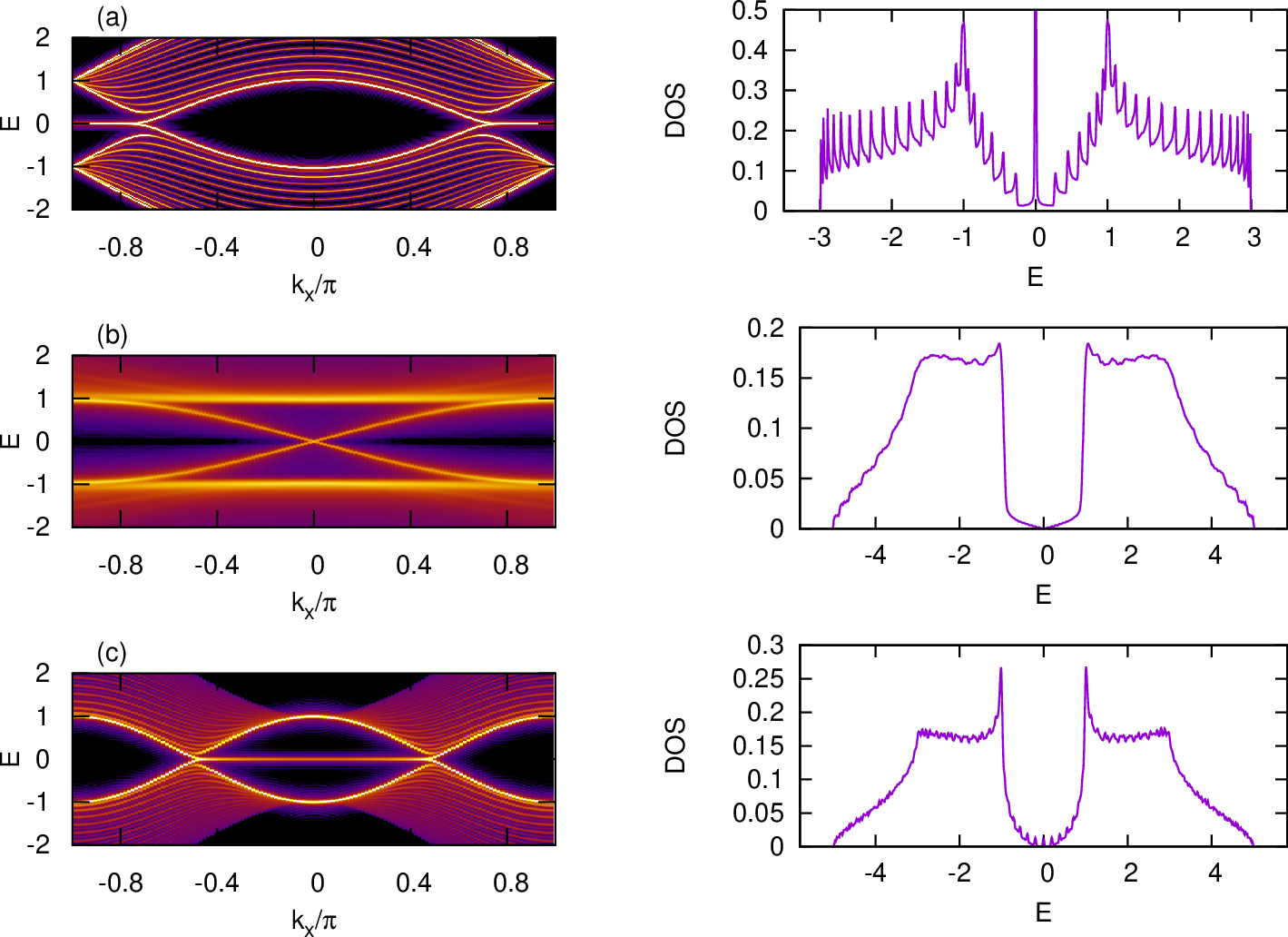}
  \caption{Band dispersion along $k_x$  deduced from the single-particle spectral function \hf{$A({\bf k},E)$}  [left panels] and density of states (DOS) [right panels] for the  model Hamiltonians~\GHcomm{\eqref{graphene_model},   \eqref{eq:TB_H_topi}, and~\eqref{eq:TB_H_wsm}}. (a) Zigzag GNR with $v_{n,j}=0$ having 16 ``rows''  and open boundary conditions (BCs)  in $y$ direction (periodic BCs  in $x$ direction).  (b) Strong TI with  $m=2$, $\Delta_{1/2} = V_n=0$ on a  cuboid with $512\times 64\times 8$ sites and periodic BCs.  \GHcomm{Here the Dirac cone (linear dispersion) near $E=0$ is due to the surface states.} (c) Weyl semimetal on a cuboid with $256\times 32\times 32$ sites and open BCs in $z$ direction (periodic BCs in $x$ and $y$ directions).
     }
  \label{fig:dos_TM_clean}
\end{figure}

The minimal theoretical models for topological Weyl semimetals have been reviewed quite recently by \cite{Cormick_weyl}. Here we consider the following  3D lattice \hf{\emph{two-band}} Hamiltonian,
\begin{eqnarray}
  H  & = &  \sum_{n=1}^{N} \left(   \Psi_{n+\hat {\bf e}_x}^\dagger \frac{\sigma_x} 2 \Psi_{n} \phantom{\sum_i^{N^4}}\right.\nonumber\\
   & & \left. {} + \sum_{j=y,z} \Psi_{n+\hat \ee_j}^\dagger \frac {\sigma_x + \ii \sigma_j} 2 
                     \Psi_{n} + \text{H.c.} \right)   \nonumber\\ % && 
           & &   + \sum_{n=1}^{N} 
              \Psi_{n}^\dagger \left[ V_n -  \sigma_x ( 2+\cos k_0 ) \right]
                    {\Psi_{n}}\,,  
              \label{eq:TB_H_wsm}
\end{eqnarray}
where $\Psi_n$  is now a two-component spinor and $\sigma_j$ are the Pauli matrices (again, the lattice constant is set to unity, just as the transfer element). In momentum space [${\bf k}=(k_x,k_y,k_z)$], the \hf{($2\times2$)   band matrix} takes the form  ($V_n=0$)
\begin{eqnarray}
  \varepsilon({\bf k}) &  =  & \sigma_x ( \cos k_x - \cos k_0 + \cos k_y + \cos k_z - 2 ) \nonumber\\
  & & {} + \sigma_y \sin k_y + \sigma_z \sin k_z\,,
                       \label{eq:band_h_wsm}
\end{eqnarray}
developing two Weyl nodes at momenta ${\bf k}^W_{\pm} = (\pm k_0, 0, 0)$ with $k_0=\pi/2$, as seen in Fig.~\ref{fig:dos_TM_clean}~(c) [\cite{Hasan:2017}].

\section{Matrix-free code for topological systems with  Dirac cones}

In general, topological materials have a rather complex lattice structure,
although it is so regular that a matrix-free formulation
of sparse matrix-vector multiplication (spMVM) and similar kernels is feasible due to the stencil-like
neighbor relations.
The lattice is always periodic (apart from disorder effects), but
particle transfer integrals or interactions vary widely among materials. In other words,
the resulting stencil geometry depends strongly on the physics, 
and in a way that makes it impossible to set up optimal code
for all possible situations in advance although the core algorithm
is always a stencil-like update scheme. The 
required blocking strategies for optimal code performance
also vary with the stencil shape. Consequently, it is worthwhile
to generate the code for a particular physical setup.
This allows to hard-code performance-relevant features and
takes out a lot of uncertainty about compiler optimizations.
In this section we describe some of the design goals and
implementation details of our matrix-free code, including
the DSL, representative benchmark cases, and performance results.

\subsection{Preliminary considerations}

%- die Systemgrößen sind typischer weise so, dass die Dauten im Hauptspeicher ligen
%- folglich ist der Datentansfer zum Hauptspeicher der Flaschenhals 

Many numerical algorithms that can describe quantum systems, such as
eigenvalue solvers or methods for computing spectral properties, such
as the Kernel Polynomial Method (KPM) (reviewed in \cite{kpm}),
require the multiplication
of large sparse matrices with one or more RHS vectors as a
time-con\-suming component. If the matrix is stored explicitly in memory
and special structures such as symmetry and dense subblocks are not
exploited, The data transfer between the CPU and the main memory is
the performance-limiting bottleneck. An upper limit for the performance
of a typical linear algebra building block such as spMVM can thus
be easily calculated by means of the naive roof{}line model, which
was popularized by \cite{roofline:2009} \gh{and applied in many different
contexts, including stencils (\cite{Datta:2009}), fluid dynamics
(\cite{Randles:2013}), and sparse linear algebra (\cite{Gropp:1999}),
among many others}:
\begin{equation}\label{eq:roofline}
  P\leq \min\left(P_\mathrm{peak},b_\mathrm S/B_\mathrm c\right)\eos
\end{equation}
  This model assumes that the performance of a loop is either limited
  by the computational peak performance of the CPU ($P_\mathrm{peak}$)
  or by the maximum performance allowed by memory data transfers
  ($b_\mathrm S/B_\mathrm c$), whichever is more stringent. In case of
  spMVM and similar algorithms on any modern multicore CPU, the former
  is much larger than the latter, so we can safely ignore it here.
  $b_\mathrm S$ is the achievable main memory bandwidth in bytes/s; it
  can be determined by a suitable benchmark, such as STREAM by \cite{stream}.
  $B_\mathrm c$ is the \emph{code balance}, i.e., the ratio of the
  required data volume through the memory interface (in bytes) and the amount of
  work (usually floating-point operations, but any valid ``work'' metric
  will do). Clearly, $b_\mathrm S/B_\mathrm c$ is then an 
  upper limit for the expected performance of the loop.  In practice
  one can determine the code balance by code inspection and
  an analysis of data access locality. % \cite{Hager:2010}.
  Whenever the data traffic
  cannot be calculated accurately, e.g., because some indirect
  and unpredictable access is involved, it is often possible to
  give at least a lower limit for $B_\mathrm c$ and thus an absolute upper
  limit for the performance. A ``good'' code in the context of the roof{}line
  model is a code whose performance is near the limit. Once this
  has been achieved, any optimization
  that lowers $B_\mathrm c$ will increase the performance accordingly.
  Many refinements of the model have been developed to make it more
  accurate in situations where the bottleneck is not so clearly
  identified, e.g., by \cite{Loft} and \cite{sthw15}.

  \gh{Depending on the processor architecture, SIMD vectorization, i.e.,
    using data-parallel instructions to carry out multiple operations
    in parallel on short vectors, may be required to achieve memory bandwidth
    saturation even with a code that has a rather low code balance.
    This happens when the single core is too slow with scalar, i.e.,
    non-SIMD code, so that even the combined demand of all cores for
    data does not exert enough ``pressure'' on the memory interface.
    Hence, SIMD vectorization was given special attention in
    our framework. See below for details.}

It was first shown by \cite{Gropp:1999} that the minimal code balance of spMVM for double precision,
real matrices in CRS format and a 32-bit index is 
$6\,\BF$, leading to memory-bound execution if the
matrix does not fit into a cache. A matrix-free formulation can greatly
reduce the demand for data and leads, in case of many
topological materials, to stencil-like update schemes.
\gh{Even if some of the terms in the operators require
  variable coefficients, getting rid of the matrix data
  still has a notable effect.} Although the resulting code
is limited by memory bandwidth as well, the code balance can be
very low depending on the particular stencil shape and
on whether layer conditions (LCs) are 
satisfied. The concept of layer conditions was conceived by
\cite{Rivera:2000} and applied in the context of advanced
analytic performance models by \cite{sthw15}. In the following we briefly 
describe the optimizations that were taken into account, using 
a simple five-point stencil as an example.

\begin{lstlisting}[caption={Two-dimensional five-point stencil sweep with one RHS and one LHS vector. The highlighted elements must come from cache for optimal code balance.},label=lst:jacobi2d,float=tbp,numbers=none,numberstyle=\tiny,belowcaptionskip=\smallskipamount]
double *x, *y; // RHS/LHS vector data
int imax,kmax; // grid size [0:imax]x[0:kmax]

for(int k=1; k<kmax; ++k) 
 for(int i=1; i<imax; ++i)
  y[i+k*(imax+1)] = 
    c1*%x[(i+1)+k*(imax+1)]% + c2*%x[(i-1)+k*(imax+1)]%
  + c3*x[i+(k+1)*(imax+1)] + c4*%x[i+(k-1)*(imax+1)]%
  + noise[i+k*(imax+1)];
\end{lstlisting}
Listing \ref{lst:jacobi2d} shows one update sweep of this code, i.e.,
one complete update of one LHS vector.  In a matrix-bound formulation
the coefficients \verb.c1., \ldots, \verb.c4.  would be stored in
memory as separate, explicit arrays. In addition to the RHS vector the array
\verb.noise[]. is read, implementing a random potential. As is 
customary with stencil algorithms we use the lattice site update
(LUP) as the principal unit of work, which allows us to
decouple the analysis from the actual number of \flops\ executed
in the loop body.

The minimum code balance of this loop nest for data in memory is
$B_\mathrm c=\left(16+8+8\right)\,\BL=32\,\BL$, because each LHS
element must be updated in memory (16\,\bytes), and each RHS and noise
element must be loaded (eight \bytes\ each). If nontemporal stores can
be used for \verb.y[]., the code balance reduces to $24\,\BL$ because
the write-allocate transfers do not occur and data is written to
memory directly. The minimum code balance can only be attained,
however, if the highlighted RHS elements do not have to be loaded from
memory. This LC is satisfied if at least three
successive rows of \verb.x[]. fit into the cache. Assuming that the 
\verb.noise[]. array and the LHS also require one row of cache
space, the condition reads:
\bq\label{eq:layercond}
5\times\mathtt{imax}\times 8~\bytes < C\cma
\eq
where $C$ is the available cache size in bytes per thread. 
In multi-threaded execution with outer loop parallelization via
OpenMP, each thread must have its LC fulfilled.
If the condition is broken,
the inner loop can be blocked with a block size of \verb.ib. and
the condition will be satisfied if 
\bq\label{eq:ibcond}
\mathtt{ib}<\frac{C}{40\,\bytes}\eos
\eq
If the blocking is done for a cache level that is shared 
among the threads in the team, the LC gets more
and more stringent the more threads are used.
For best single-threaded performance it is advisable to block for
an inner cache, i.e., L1 or L2\@. See \cite{sthw15} for more details.
Our production code determines the optimal block size according
to (\ref{eq:ibcond}).

The \verb.noise[]. array is a significant contribution to the 
code balance. However, its contents are static and can be generated
on the fly, trading arithmetic effort for memory traffic. 
The generation of random numbers should certainly be fast 
and vectorizable, so as to not cause too much overhead.
See the section on random number generation below for details.

\begin{lstlisting}[caption={Two-dimensional five-point stencil sweep with \texttt{r} RHS and LHS vectors. SIMD vectorization across RHS and LHS vectors is possible and efficient if the vector storage order can be chosen as shown.},label=lst:jacobi2d-r,float=tbp,numbers=none,numberstyle=\tiny,belowcaptionskip=\smallskipamount]
double *x, *y; // RHS/LHS vector data
int imax,kmax; // grid size [0:imax]x[0:kmax]

for(int k=1; k<kmax; ++k) 
 for(int i=1; i<imax; ++i)
  %for(int s=0; s<nb; ++s)%
   y[%s%+%nb%*i+%nb%*k*(imax+1)] = 
     c1*x[%s%+%nb%*(i+1)+%nb%*k*(imax+1)] 
   + c2*x[%s%+%nb%*(i-1)+%nb%*k*(imax+1)]
   + c3*x[%s%+%nb%*i+%nb%*(k+1)*(imax+1)] 
   + c4*x[%s%+%nb%*i+%nb%*(k-1)*(imax+1)]
   + noise[%s%+%nb%*i+%nb%*k*(imax+1)];
\end{lstlisting}Some algorithmic variants require the concurrent, independent
execution of stencil updates on multiple source and target vectors.
Although SIMD vectorization is easily possible even with a single
update by leveraging the data parallelism along the inner dimension,
a more efficient option exists for multiple concurrent updates:
If the vectors can be stored in an interleaved way, i.e., with the
leading dimension going across vectors, vectorization along 
this dimension is straightforward if the number of vectors
is large compared to the SIMD width. As opposed to the traditional
scheme, perfect data alignment can be achieved (if this is required)
and no shuffling of data in SIMD registers is necessary for optimal
register reuse. See Listing~\ref{lst:jacobi2d-r} for an example using
a simple five-point stencil. The considerations about LCs
do not change apart from the fact that now each RHS vector needs to have
its own LC fulfilled. Condition (\ref{eq:ibcond}) is thus modified
to
\bq\label{eq:ibcond-r}
\texttt{ib}<\frac{C}{5\times n_b\times 8\,\bytes}
\eq
if \verb.n_b. is the number of concurrent updates and the \verb.noise[]. arrays
have to be loaded from memory.

\gh{In summary, code generation in our framework has two goals:
  produce a spMVM routine with minimal code balance by cache-adapted
  loop blocking, and produce SIMD-vectorized code in order to be able
  to address the memory bandwidth bottleneck in all relevant cases.}

%\subsection{Implementation}
%
% - Eine generalisierte Stencildarstelung wurde entwickelt, die viele Problemf\"alle (auch aus\ss halb der TM) abdeckt. \\
% - Ein aufwendiges Precompiler-Konzept in Form eines Python-Script wurde entwickelt,
%   welches aus der  generalisierte Stencildarstelung effizenten und multiplen Kernels f\"ur
%   Unterschiedliche Anforderungen erstellt. \\
% - Vektorblocking erhöht zuzätzlich die SIMD-Effizenz (auch bei gleicher Code-Ballance) \\
% - Problemf\"all sind (generell) Memory-Bound  \\
% - Loop Fusion (z.B. partial dotpoducts on the fly) und Cache-Blockng 
%

\subsection{Domain-Specific Language (DSL)}

In order to provide a maximum amount of flexibility to users and still
guarantee optimal code, a DSL was constructed which is used to define
the physical problem at hand. A precompiler written in Python then
generates OpenMP-parallel C code for the sparse matrix-vector multiplication
(a ``lattice sweep''), which can be handled by a standard
compiler. In the following we describe the DSL in detail by example. 

The source code for the DSL program resides in a text file. The code
begins with a specification of the problem dimensionality (2D/3D) 
and the basis size:
\begin{lstlisting}
  dim 2
  size 4
\end{lstlisting}
The stencil coefficients can take various forms: \emph{constant},
\emph{variable}, or \emph{random}. The number of coefficients 
of each kind is set by the keywords \verb.n_coeff_*., where 
``\verb.*.'' is one of the three options. For example, in case of four 
variable coefficients:
\begin{lstlisting}
  n_coeff_variable 4
\end{lstlisting}
The command \verb.nn. with two or three arguments (depending on the
\verb.dim. parameter) and the following \verb.size. lines define a sparse
coefficient matrix to a neighboring lattice block at the offset
defined by the arguments of \verb.nn.\@.
Multiple entries in a line are separated by ``\verb.;.''. 
Optionally the first entry begins with ``\verb.l.'' followed by the row index.
A single block entry is written as ``\verb.{column index}|{value}.'' for
a fixed value, or as
``\verb.{column index}|. \verb.{type}|{type index or value}.'' for a different type.
This is a simple example for a coefficient matrix one lattice position to the
left of the current position (set by \verb.nn -1 0.) and a fixed entry of value $-1$ at
position (0,1) and another entry of value $-1$ at position (3,2):
\begin{lstlisting}
nn -1 0 
l0;        1|-1       
l1;                   
l2;                   
l3;               2|-1       
\end{lstlisting}
Note that all indexing is zero-based. 
The following coefficient types are allowed:
\begin{itemize}
\item[\texttt{f}] Fixed coefficient, hard-coded (default type). This means that
  an entry
  of ``\verb.1|-1.'' can also be written as ``\verb.1|f|-1.''\@. It will be
  hard-coded into the generated C code.
\item[\texttt{c}] Constant coefficient per lattice site, read from an array of
  length \verb.n_coeff_const.\@. For example, ``\verb.1|c|2.'' means that the coefficient
  used will be \verb.coeff_c[2].\@. The coefficient array can be changed at
  runtime if required, or preset in the DSL source code. For example, the line
  \begin{lstlisting}
  coeff_const_default 1 2 -0.5
  \end{lstlisting}
  will initialize the array \verb.coeff_c[]. with the specified values.
\item[\texttt{v}] Variable coefficient per lattice site, read from an array of length
  \verb.n_coeff_variable. per lattice site. 
\item[\texttt{r}] Random coefficient per lattice site, read from an array of length
  \verb.n_coeff_rand. per lattice site. 
\end{itemize}
% - The command ``nn \textless $dx$\textgreater\ \textless $dy$\textgreater\ \textless $dz$\textgreater'' and the following $S$ lines define
%      a spare coefficient matrix to a neighbouring lattice block. \\
% - The $n$the line contains the entries for the $n$the subsize separated by ``;''. \\
% - The information for each entry is separated by ``\textbar''.
%     The first string is the subsize index of the neighbouring lattice block and the second string is the value.
%     If the value is not fixed, there can use the character ``c'' for a constant, ``v'' for a variabel or ``r'' for a random coefficent. 
%     Then third string is the index of the non-fixed coefficent kind. \\
% - The number of the non-fixed coefficents is seted by ``n\_coeff\_variable \textless $nc$\textgreater'',
%       ``n\_coeff\_variable \textless $nv$\textgreater'' and ``n\_coeff\_rand \textless $nr$\textgreater''. \\
% - A random coefficent can reused only in the same lattice block.

In Listing~\ref{lst:graphenedsl} we show a complete example for a 2D graphene
stencil with variable coefficients on the diagonal, while
Listing~\ref{lst:graphenec} shows the generated C source code for the
spMVM. \gh{Note that spatial
  blocking is the only explicit optimization done by the code generator.
  We rely on the compiler to produce SIMD-vectorized code from the
  C source, which is usually possible by giving it sufficient
  information, particularly about non-aliasing of pointers}.\footnote{\gh{This can either be achieved by using the \texttt{restrict} keyword on pointer declarations, by compiler directives such as \texttt{ivdep}, or via global options such as \texttt{-fno-alias}}.}
\begin{lstlisting}[float=tbp,caption={DSL source for a graphene stencil},numbers=none,label=lst:graphenedsl,multicols=2,basicstyle=\color{red}\footnotesize\ttfamily]
dim 2
size 4
n_coeff_variable 4

nn -1 0 
l0;   1|-1       
l1;              
l2;              
l3;         2|-1       

nn 0 -1
l0;              3|-1
l1;      
l2;
l3;




nn 0 0
l0; 0|v|0; 1|-1              
l1; 0|-1;  1|v|1; 2|-1       
l2; 1|-1;  2|v|2; 3|-1
l3;       2|-1;  3|v|3

nn 0 1
l0; 
l1; 
l2;              
l3; 0|-1       

nn 1 0
l0;      
l1; 0|-1                     
l2;               3|-1
l3;                           
\end{lstlisting}
\begin{lstlisting}[float=tbp,caption={Generated matrix-vector multiplication code for the graphene stencil (shortened)},style=customc,label=lst:graphenec,basicstyle=\color{red}\ttfamily\footnotesize]
#pragma omp for schedule(static,1) nowait
for(int i2=sys->b_y; i2<sys->n_y+sys->b_y; i2++){ // y loop
 for(int i1=i1_0; i1<i1_end; i1++){ // z loop
  int j = i1 +  ldz * ( i2 + ldy * i3);
  int i  = 4 * j;
  #pragma vector aligned
  for(int k=0; k<4; k++) {  // vector block loop
   y[(i+0)*4+k] =  scale_z[k] * y[(i+0)*4+k] + scale_h[k] 
    * ( -shift_h[k] * x[ (i+0)*4+k ]
    -1. * x[(i+1+4*(0+ldz*(-1)))*4+k]
    -1. * x[(i+3+4*(-1+ldz*(0)))*4+k]
    -1. * x[(i+1+4*(0+ldz*(0))*4+k]
    +coeff_v[j*ldv+0] * x[(i+0+4*(0+ldz*(0)))*4+k]);
   y[(i+1)*4+k] = ...;
   y[(i+2)*4+k] = ...;
   y[(i+3)*4+k] = ...;
  }}
  if( p->dot_xx || p->dot_xy || p->dot_yy ) {
  for(int i1=i1_0; i1<i1_end; i1++) {
   int j = i1 +  ldz * ( i2 + ldy * i3);
   int i  = 4 * j;
   #pragma  vector aligned
   for(int k=0; k<4; k++) {
    xx[k]+=x[(i+0)*4+k]*x[(i+0)*4+k]+x[(i+1)*4+k]*x[(i+1)*4+k]+x[(i+2)*4+k]*x[(i+2)*4+k]+x[(i+3)*4+k]*x[(i+3)*4+k];
    xy[k]+=...;
    yy[k]+=...;
}}}}
\end{lstlisting}
%% GH: Code, der von oben geloescht wurde
%%   y[(i+1)*4+k] =  scale_z[k] * y[(i+1)*4+k] + scale_h[k] 
%%    * ( -shift_h[k] * x[ (i+1)*4 + k ]
%%    -1. * x[(i+0+4*(0+ldz*(0)))*4+k]
%%    -1. * x[(i+2+4*(0+ldz*(0)))*4+k]
%%    +coeff_v[j*ldv+1] * x[(i+1+4*(0+ldz*(0)))*4+k]
%%    -1. * x[(i+0+4*(0+ldz*(1)))*4+k]);
%%   y[(i+2)*4+k] =  scale_z[k] * y[(i+2)*4+k] + scale_h[k] 
%%    * ( -shift_h[k] * x[(i+2)*4+k]
%%    -1. * x[(i+1+4*(0+ldz*(0)))*4+k]
%%    -1. * x[(i+3+4*(0+ldz*(0)))*4+k]
%%    +coeff_v[j*ldv+2] * x[(i+2+4*(0+ldz*(0)))*4+k]
%%    -1. * x[(i+3+4*(0+ldz*(1)))*4+k] );
%%   y[(i+3)*4+k] =  scale_z[k] * y[(i+3)*4+k] + scale_h[k] 
%%    * ( -shift_h[k] * x[(i+3)*4+k]
%%    -1. * x[(i+2+4*(0+ldz*(-1)))*4+k]
%%    -1. * x[(i+2+4*(0+ldz*(0)))*4+k]
%%    +coeff_v[j*ldv+3] * x[(i+3+4*(0+ldz*(0)))*4+k]
%%    -1. * x[(i+0+4*(1+ldz*(0)))*4+k] );
%%
%%xy[k]+=x[(i+0)*4+k]*y[(i+0)*4+k]+x[(i+1)*4+k]*y[(i+1)*4+k]+x[(i+2)*4+k]*y[(i+2)*4+k]+x[(i+3)*4+k]*y[(i+3)*4+k];
%%    yy[k]+=y[(i+0)*4+k]*y[(i+0)*4+k]+y[(i+1)*4+k]*y[(i+1)*4+k]+y[(i+2)*4+k]*y[(i+2)*4+k]+y[(i+3)*4+k]*y[(i+3)*4+k];
Coefficient arrays and execution parameters (such as, e.g., the grid size)
can be configured and changed at runtime.
The code repository\footnote{\url{http://tiny.cc/PVSC-DTM}} contains
numerous examples that demonstrate the DSL usage and how the generated
source code can be embedded into algorithms.

\subsection{Random number generator}\label{sec:rng}

\begin{lstlisting}[float=tbp,caption={Using an ``out-of-band'' fast RNG to save memory data traffic},label=lst:oobrng]
for(int k=1; k<kmax; ++k) {
  for(int i=1; i<imax; ++i)
    random[i] = ...; // fast RNG
  for(int i=1; i<imax; ++i)
    y[i+k*(imax+1)] = 
      c1*x[(i+1)+k*(imax+1)] + c2*x[(i-1)+k*(imax+1)]
    + c3*x[i+(k+1)*(imax+1)] + c4*x[i+(k-1)*(imax+1)]
    + random[i];
}
\end{lstlisting}
In some physically relevant cases, the Hamiltonian matrix has
a diagonal, random component. These random numbers are usually
stored in a constant array to be loaded during the matrix-vector
multiplication step. At double precision this leads to an increase
in code balance by 8\,\BL, which can be entirely saved by
generating the random numbers on the fly using a fast random
number generator (RNG)\@. Considering that the stencil update
schemes studied here can run at several billions of lattice
site updates per second on modern server processors, a suitable
RNG must be able to produce random numbers at a comparable
rate. This is not possible with standard library-based implementations
such as, e.g., \verb.drand48()., but faster and (in terms
of quality) better options do exist. The RNG code should
be inlined with the actual spMVM or at least be available
as a function that generates long sequences of random numbers
in order to avoid hazardous call overhead.

The standard type of RNG used in scientific computing is the linear
congruential generator (LCG), which calculates $ x_{i+1} = (a x_i + b)
\mod m $ and casts the result to a floating-point number. The numbers
$a$, $b$, and $m$ parameterize the generator; for efficiency reasons
one can choose $m$ to be a power of two (e.g., $m=2^{48}$ in
\verb.drand48().), but such simple methods fail the statistical tests
of the popular TESTU01 suite devised by \cite{TestU01}\@. However, if there are
no particular quality requirements (i.e., if only ``some randomness''
is asked for), they may still be of value. 
Despite the  nonresolvable dependency of
$x_{i+1}$ on $x_i$, which appears to rule out SIMD vectorization,
LCGs can be vectorized if a number of independent random number
sequences is needed and if the SIMD instruction set of the hardware
allows for the relevant operations (e.g., SIMD-parallel addition,
multiplication, and modulo on unsigned integer operands with the
required number of bits)\@.

A more modern and similarly fast approach to RNGs are the
\emph{xorshift} generators by  \cite{Xorshift_2003}.
In the simplest case they work by a sequence of XOR mask operations of
the seed with a bit-shifted version of itself: $x\ \char`\^{=}\: x \ll
a; x\ \char`\^{=}\: x \gg b; x\ \char`\^{=}\: x \ll c$\@. Improved versions
like the \emph{xorshift128+} by \cite{Xorshift128plus_2014} pass all
statistical tests of the ``SmallCrush'' suite in TESTU01\@.
Table~\ref{tab:rngcomp} shows a performance comparison of different
RNGs on one socket of the IVB and BDW systems, respectively.
\begin{table*}[tb]
\centering
\begin{tabular}{ l c>{\centering\arraybackslash}m{2cm}>{\centering\arraybackslash}m{2cm}}
  RNG  (loop unrolling) & failed SmallCrush &  Perf.~IVB [\GRNS]&  Perf.~BDW [\GRNS]\\ \hline
  lgc32 (32)           & 1,2,3,4,5,6,7,8,9,10   &  19.3 & 28.8 \\
  xorshift32 (32)      & 1,3,4,8,9,10           &  10.1 & 28.8 \\
  xorshift128 (16)      & 6                      &   8.29 & 25.4 \\
  xorshift64\_long (16)   & 8                    &   8.26 & 31.7 \\
  xorshift128plus\_long (16)  &          ---     &   6.34 & 20.7 \\
  Intel MKL SFMT19937  &          ---     &   7.72 & 22.0 \\
\end{tabular}
\caption{\label{tab:rngcomp}Performance comparison of the LCG32 RNG,
  four xorshift generators of different
  sophistication, and the SFMT19937 generator available in the Intel MKL.
  Performance numbers are given in billions of random
  numbers per second (\GRNS) on one socket (10 or 18 cores,
  respectively).  The benchmark consists in the (repeated)
  computation of 2000
  double-precision random numbers with uniform distribution.
  For each generator, the
  second column lists the failed SmallCrush tests of the TESTU01 suite.
  These particular test results were taken with the Intel C
    compiler version 17.0 update 5. The benchmarks are available in
    the code repository.}
\end{table*}
The speedup between IVB and BDW is particularly large for the
xorshift generators because the AVX2 instruction set on BDW supports
SIMD-parallel bit shifting operations, which are not available
in AVX\@. For reference, we have included a SIMD-vectorized
  Mersenne Twister RNG (SFMT19937), which is available in Intel's
  Math Kernel Library (MKL). \gh{Note that the purpose of
    Table~\ref{tab:rngcomp} is not to compare the performance
    of different RNG algorithms but to give an impression of how
    expensive random number generation is compared to the pure
    spMVM operation. To this end, the GRN/sec numbers should be
    compared to the GLUP/s stencil performance numbers in the following
    sections}.

Whether a particular RNG impacts the performance of the matrix-free
spMVM step depends on the application. In the cases we investigate
here, a whole row of random numbers can be generated in advance before
each inner loop traversal (see Listing~\ref{lst:oobrng}) without a
significant performance degradation in the bandwidth-saturated case.
Fusing the RNG with the inner loop is \gh{a possible optimization
  that would, however, not change the roof{}line limits but
  at best lead to faster bandwidth saturation as the number
  of cores goes up}.

\subsection{Geometry}

The current implementation of \pvsctm\ supports cuboid domains
of arbitrary size (only limited by memory capacity) with 3D
domain decomposition using MPI, and OpenMP-based multithreading on
the subdomain level. For each spatial dimension, periodic
BCs can be configured separately as needed. 
Variable-coefficient arrays and initial vectors can be
preset via a user-defined callback function
with the following interface:
\begin{lstlisting}
  double (* fnc)(long * r, int k, \
                 void * parm , void * seed);
\end{lstlisting}
This function must expect the following input parameters:
\begin{lstlisting}
  { x=r[0], y=r[1], z=r[2], \
    basis_place=r[3], vector_block_index=k }
\end{lstlisting}
It returns the respective vector entry as a double-precision number.
% hier fehlt noch eine Beschreibung der Bedeutung der versch. Eintraege
The pointer \verb.parm. is handed down from the vector initial call
and allows for configuring specific options. The pointer
\verb.seed. is a reference to a 128-bit process- and thread-local
random seed.

Finally a vector block will be initialized by calling the function:
\begin{lstlisting}
pvsc_vector_from_func( pvsc_vector * vec, \
      pvsc_vector_func_ptr * fnc, void * parm);
\end{lstlisting}
This mechanism lets the user define a generalized initial function
with optionally free parameters. In addition, a thread-local random
seed for optional random functions is available in the initialization
function, which enables a fully parallelized initialization of
vectors.

\subsection{Benchmarks}

In order to validate the performance claims of our matrix-free
implementation and optimization of random number generation
we ran several test cases on the benchmark systems described
in the introduction. Performance is quantified in billions of lattice
site updates per second (\GLUPS)\@. For all stencil update
kernels (spMVMs) with constant coefficients studied here, the minimal code balance is
24\,\BL\ with on-the-fly RNGs \GHcomm{(see Listing~\ref{lst:oobrng})}
and 32\,\BL\ with random numbers
stored as constant arrays. The roof\/line model thus predicts
bandwidth-bound per-socket upper performance limits of
1.67\,\GLUPS\ on IVB and 2.5\,\GLUPS\ on BDW\@.
\begin{figure}%[htbp]
  \centering\includegraphics[width=0.9\linewidth,clip]{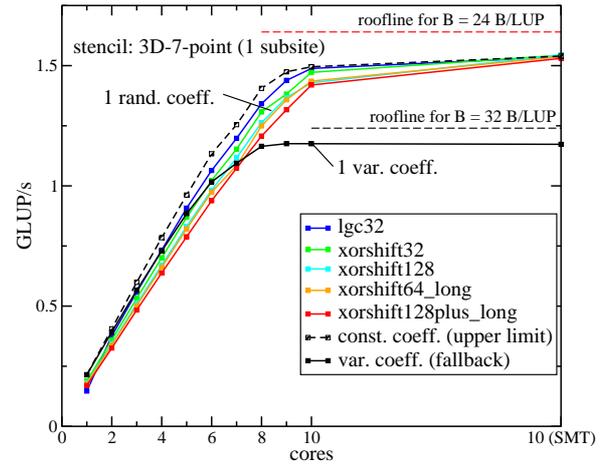}
  \caption{Performance scaling of \GHcomm{a spMVM kernel using
    a} constant-coefficient 3D 7-point
    stencil problem with \GHcomm{(as in Listing~\ref{lst:oobrng})}
    and without on-the-fly RNG on one IVB socket (10
    cores) and with SMT (2 threads per core)\@.  The dashed line shows
    the performance without a random potential, whereas the filled
    black squares (fallback) show the result with random numbers read
    from memory. All other data sets were obtained with different
    on-the-fly RNGs ($n_b = 1$, system size $512^3$).  }
  \label{fig:rand_on_the_fly_ivb}
\end{figure}

\begin{figure}%[htbp]
  \centering\includegraphics[width=0.9\linewidth,clip]{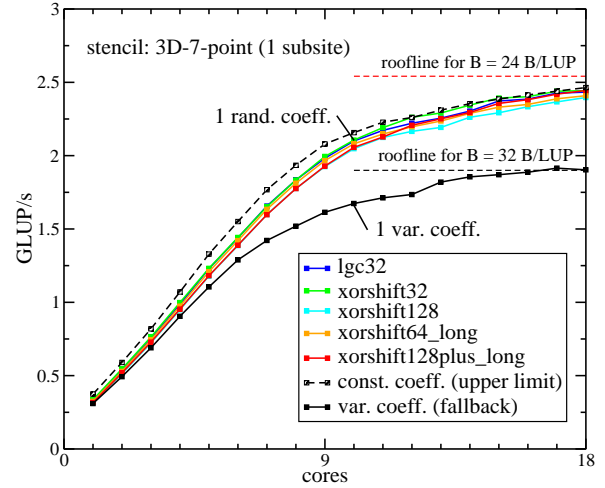}
  \caption{Performance scaling as in Fig.~\ref{fig:rand_on_the_fly_ivb} but
  on one BDW socket (18 cores)\@.}
  \label{fig:rand_on_the_fly_bw}
\end{figure}
Figures \ref{fig:rand_on_the_fly_ivb} and \ref{fig:rand_on_the_fly_bw}
show the performance scaling of the spMVM with a 3D 7-point stencil on
one socket of the benchmark systems. On IVB, the ``fallback'' kernel, which
uses explicitly stored random numbers, saturates the memory bandwidth
with eight cores at about 95\% of the achievable bandwidth
(black solid line)\@. The kernel without random numbers (labeled
``const. coeff.'') marks a practical upper performance limit. It also
saturates at about the same bandwidth (and thus at 33\% higher
performance), with a very slight additional speedup from SMT
(Hyper-Threading)\@. As expected, the versions with on-the-fly RNGs
are somewhat slower on the core level due to the increased amount of
work, which, in case of the xorshift variants, leads to
a lower performance than for the fallback variant up to seven
cores, and non-saturation
when only physical cores are used. SMT can close this gap by filling
pipeline bubbles on the core level, and all RNG versions end up
at exactly the same performance with 20 SMT threads.
On BDW the full-socket situation is similar, but all versions come
closer to the practical bandwidth limit than on IVB, and the fallback
variant is slower than all RNG versions at all core counts.

The bottom line is that even the most ``expensive'' on-the-fly RNG
allows memory bandwidth saturation on both architectures, that the
roof\/line model predictions are quite accurate, that the automatic
spatial blocking in \pvsctm\ works as intended and yields the optimal
in-memory code balance, and that the elimination of the stored random
number stream causes the expected speedup even with high-quality RNGs.
\begin{figure}%[htbp]
  \centering\includegraphics[width=0.9\linewidth,clip]{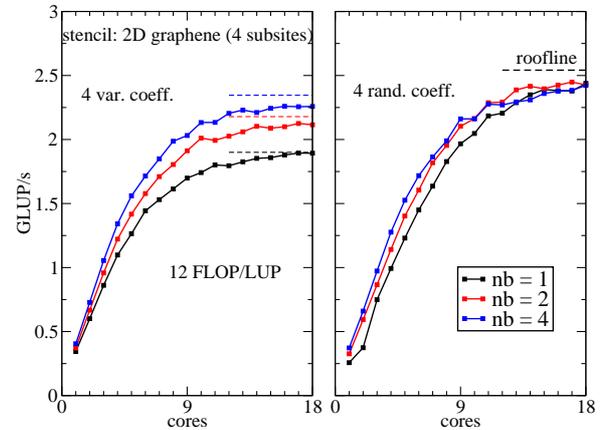}
  \caption{Performance scaling of \GHcomm{spMVM with} graphene stencil kernels
    \GHcomm{on blocks of vectors of size $n_b=1$, 2, and 4} with
    variable coefficients
    (left) and on-the-fly RNGs (right) on BDW
    (system size $8000 \times 8000/n_b$).  }
  \label{fig:rand_on_the_fly_graphene_bw}
\end{figure}

Figure \ref{fig:rand_on_the_fly_graphene_bw} shows a performance
comparison of stored random numbers and on-the-fly RNG for a 2D
graphene
application with four subsites, a block vector size
$n_b$ of 1, 2, and 4, and four variable coefficients.  The code
balance goes down from 32\,\BL\ to 28\,\BL\ and finally to
26\,\BL\ when going from $n_b=1$ to 2 and 4, approaching the limit of
24\,\BL\ at $n_b\to\infty$\@. With on-the-fly RNGs substituting the
variable-coefficient arrays this balance is
achieved for any $n_b$, which is shown in the right panel of
Fig.~\ref{fig:rand_on_the_fly_graphene_bw}\@. It can also be observed
in the data that the improved SIMD vectorization with $n_b>1$
speeds up the code measurably in the nonsaturated regime,
but this advantage vanishes close to saturation because the
data transfer becomes the only bottleneck.
\begin{figure}%[htbp]
  \centering\includegraphics[width=0.9\linewidth,clip]{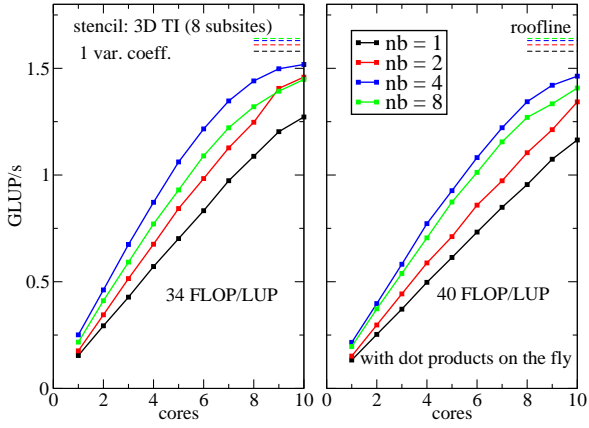}
  \caption{Performance scaling of the stencil \GHcomm{spMVM}
    kernel of a 3D TI model without (left) and with (right)
    on-the-fly dot products for $n_b=1$, 2, 4, and 8 on
    IVB\@.  (system size $256^2\times 256/n_b$)
           %Performance measurement on a Intel\textsuperscript{\textregistered} Xeon\textsuperscript{\textregistered} 2660v2 CPU via LIKWID\cite{likwid} by fixed clockspeed 2.2 GHz.
           }
  \label{fig:performance_topi_ib}
\end{figure}

\begin{figure}%[htbp]
  \centering\includegraphics[width=0.9\linewidth,clip]{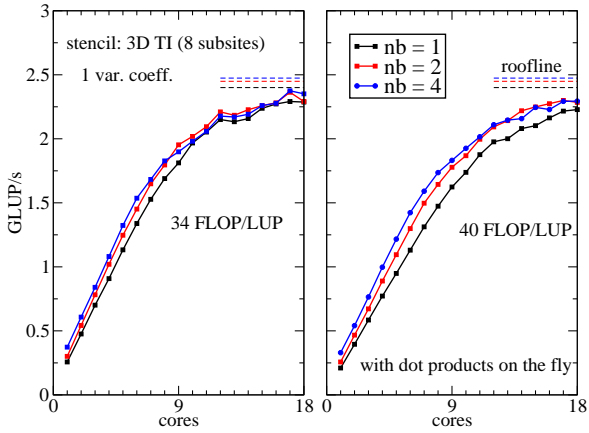}
  \caption{Performance scaling of the stencil \GHcomm{spMVM}
    kernel of a 3D TI model without (left) and with (right)
    on-the-fly dot products for $n_b=1$, 2, and 4 on
    BDW. (system size $256^2\times 256/n_b$)
           %Performance measurement on a Intel\textsuperscript{\textregistered} Xeon\textsuperscript{\textregistered} E5-2697 CPU via LIKWID\cite{likwid} by fixed clockspeed 2.2 GHz.
           }
  \label{fig:performance_topi_bw}
\end{figure}
Figures \ref{fig:performance_topi_ib} and
\ref{fig:performance_topi_bw} show the performance of the
stencil kernels of a 3D TI model with different $n_b$ on
IVB and BDW\@. Two versions are shown for each architecture:
The standard one and an optimized version with dot products
fused into the stencil kernel, increasing the number
of flops per update by six.
%\GHcomm{Das sollte oben eine eigene subsection bekommen?} 
The code balance for TI stencils is
lower than for graphene or the 7-point stencil, hence
more cores are required for bandwidth saturation.

At larger $n_b$ the loop body becomes more complicated,
and the benefit of SIMD vectorization may be compensated
by a more inefficient in-core execution due to register
shortage and less effective out-of-order execution.
This can be seen on IVB at $n_b=8$, where the available
number of cores is too small to reach saturation, as opposed
to $n_b=4$, where the SIMD width matches the number
of block vectors.

Calculating dot products on the fly has a rather small impact on
performance (less that 15\%), which on BDW vanishes at
saturation because of its generally lower machine balance. Still,
overall the roof\/line model provides a good estimate of the expected
socket-level performance of our matrix-free codes even for topological
insulators.  Note, however, that the saturation properties depend on
many factors, such as the number of cores per socket, the memory
bandwidth, the clock speed, and the SIMD width. An accurate prediction
of speedup versus the number of cores would require a more advanced
performance model, such as the ECM model described in~\cite{sthw15}\@.
\begin{figure}%[htbp]
  \centering\includegraphics[width=0.9\linewidth,clip]{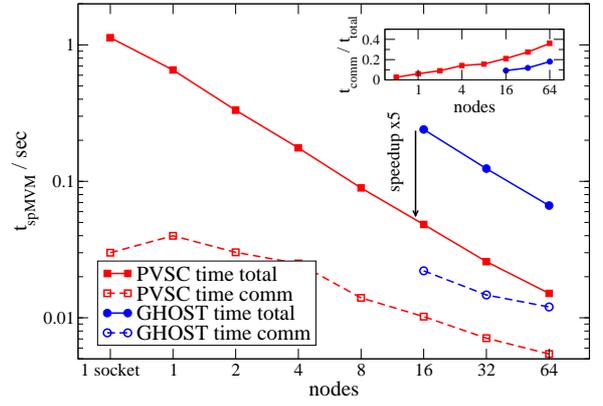}
  \caption{Runtime of spMVM ($n_b=1$) for a strong scaling test case
    of the TI model (system size $608^3$) with 200 iterations on the
    Emmy cluster, comparing \pvsctm\ with the GHOST
    library.  The dashed lines show
    the communication time only. Inset: ratio of
    communication time vs.\ total runtime. All codes were run with one
    MPI process per socket (ten cores) and ten OpenMP threads per process.}
  \label{fig:strong_scaling_nb1}
\end{figure}

\begin{figure}%[htbp]
  \centering\includegraphics[width=0.9\linewidth,clip]{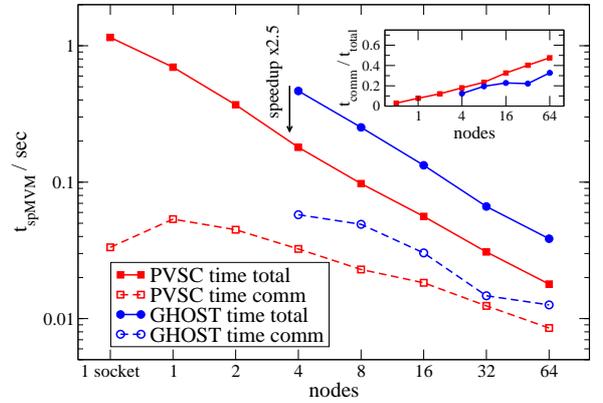}
  \caption{Runtime of spMVM
    ($n_b=4$) for a strong scaling test case
      of the TI model (system size $384^3$) with 200 iterations on the
      Emmy cluster, comparing \pvsctm\ with the GHOST
      library.  The dashed lines show the
      communication time only. Inset: ratio of communication time
      vs.\ total runtime. All codes were run with one
    MPI process per socket (ten cores) and ten OpenMP threads per process.}
  \label{fig:strong_scaling_nb4}
\end{figure}
In Figures \ref{fig:strong_scaling_nb1} and \ref{fig:strong_scaling_nb4}
we compare \pvsctm\ with the GHOST library developed by \cite{GHOST2016} using a
strong scaling TI test case on the Emmy cluster.  One MPI process was
bound to each socket, with Open\-MP parallelization (ten threads)
across the cores.
Both \pvsctm\ and GHOST were used in ``vector mode,'' i.e.,
without overlap between communication and computation.
GHOST always uses explicitly stored matrices, which is why \pvsctm\
not only has the expected performance advantage due to its matrix-free
algorithms but also requires less hardware to handle a given problem
size. The maximum number of nodes was chosen such that a maximum
communication overhead of about 40--50\% (see insets)
can be observed for \pvsctm,
which is a reasonable upper limit in production runs for resource
efficiency reasons. \gh{Note that GHOST exhibits a larger communication
  time than PVSC-DTM because it assumes a general matrix and cuts it
  into horizontal blocks, resulting in sub-optimal communication
  behavior for stencil-based patterns. Our generated code, on the
  other hand, can
  exploit the regular next-neighbor exchange pattern}.

In the test case in Fig. \ref{fig:strong_scaling_nb1}, GHOST requires
at least 16 nodes for storing the matrix and two vectors.  With the
same resources, \pvsctm\ is about 5$\times$ faster, and can outperform
GHOST already with four nodes. The ratio of communication to
computation time is naturally larger with \pvsctm\ due to the faster
code execution. Although this particular test case cannot be
run on a single node with GHOST, the performance comparison at 16
nodes also reflects quite accurately the per-node (or per-socket,
i.e., pure OpenMP) performance ratio between \pvsctm\ and GHOST,
since at this point the communication overhead is still only
10--20\%.

For $n_b>1$ the memory traffic caused by the matrix becomes less
significant and the speedup of \pvsctm\ versus GHOST gets smaller.  In the
smaller $n_b=4$ test case shown in Fig. \ref{fig:strong_scaling_nb4}
GHOST requires at least four nodes and is still about 2.5$\times$
slower than \pvsctm\ at that point. Again, this is also
the expected per-socket speedup if it were possible to run the
test case on a single socket with GHOST.

\section{Algorithms and application examples}

\gh{So far we have set the stage for the quantum physics context of
  possible applications of our framework.  We have also described the
  performance properties of generated code and shown that it achieves
  near-optimal performance for the sparse matrix-vector multiplication
  (as given by the memory bandwidth limitation and the minimal code
  balance) on two different processor architectures for operators
  relevant for real applications.  In the following sections we give
  some examples for typical applications in the field of Dirac and
  topological materials that utilize spMVM as a major numerical
  component.}

In large-scale simulations of any kind, avoiding global
synchronization points is crucial for scalability.  This challenge can
be met by modern matrix polynomial methods.  The kernel polynomial
method (KPM), the Chebyshev time propagation
approach described in~\cite{WF08a} and \cite{Alver_PhysRevB.77.045125}, and the
high-performance Chebyshev
filter diagonalization technique (ChebFD) implementation
introduced in~\cite{Pieper_ChebFD} are
already available in \pvsctm. These algorithms
benefit from partial dot products, vector blocking, and loop
fusion.  The high-order commutator-free exponential time-propagation
algorithm introduced by~\cite{cfet} for driven quantum systems will be implemented
in the near future.

\subsection{Time Propagation}
The time evolution of a quantum state $|\psi\rangle$ is 
described by the Schr{\"o}dinger equation. 
If the Hamilton operator $H$ does not explicitly depend on the time $t$ 
we can formally integrate this equation and express 
the dynamics in terms of the time evolution operator $U(t,t_0)$ as 
$|\psi(t)\rangle = U(t,t_0)|\psi(t_0)\rangle$ with $U(t,t_0) =
e^{ -i H (t-t_0)}$ ($\hbar=1$). 
Expanding the time evolution operator 
into a finite series of first-kind Chebyshev polynomials of order $k$,
$T_k(x)=\cos(k\, \mathrm{arccos}(x))$,
we obtain~[\cite{Tal-Ezer1984,FEHSKE20092182}]
\begin{equation}
  U(\Delta t) =  e^{-ib \Delta t} 
  \Big[ c_0(a\Delta t) + 2\sum\limits_{k=1}^{M} c_k(a\Delta t)
    T_k(\tilde{H}) \Big].
\label{eq:U_1}
\end{equation}
Prior to the expansion the Hamiltonian has to be shifted and rescaled such that
the spectrum of $\tilde{H} = (H-b)/a$ is within the definition interval of the
Chebyshev polynomials, $[-1,1]$, where $a$ and $b$ are calculated from the extreme eigenvalues of $H$: 
$b=\frac{1}{2}(E_{\mathrm{max}}+E_{\mathrm{min}})$ and 
$a=\frac{1}{2}(E_{\mathrm{max}}-E_{\mathrm{min}}+\epsilon)$.
The expansion coefficients $c_k$ are given by % ~\cite{FEHSKE20092182} 
\begin{equation}
  c_k(a\Delta t) = \int\limits_{-1}^1 
  \frac{T_k(x)e^{-{\rm i} x a \Delta t }}{\pi \sqrt{1-x^2}}  dx =
  (-{\rm i})^k J_k(a \Delta t)
\end{equation}
($J_k$ denotes the $k$-th order Bessel function of the first kind). 

Calculating the evolution of a state $|\psi(t_0)\rangle$ from one 
time grid point to the adjacent one,
$|\psi(t)\rangle = U(\Delta t)|\psi(t_0)\rangle$, we
have to accumulate the $c_k$-weighted vectors
$|w_k\rangle = T_k(\tilde{H})|\psi(t_0)\rangle$.
Since the coefficients $c_k(a\Delta t)$ depend on the time step but not
on time explicitly, we need to calculate them only once.
The vectors $|w_k\rangle$ can be computed iteratively,  exploiting
the recurrence relation of the Chebyshev polynomials,
$|w_{k+1}\rangle = 2\tilde{H} |w_k\rangle - |w_{k-1}\rangle$, 
with $|w_1\rangle = \tilde{H} |w_0\rangle$ and $|w_0\rangle = |\psi(t_0)\rangle$.
Evolving the wave function from one time step to the next
then requires $M$ MVMs  of a given complex vector with 
the (sparse) Hamilton matrix of dimension $N$ 
and the summation of the resulting vectors 
after an appropriate rescaling. Thus, for time-independent $H$, arbitrary large time steps are
in principle possible at the expense of increasing $M$. We may choose $M$ such that for $k>M$ the modulus of all expansion coefficients 
$|c_k(a\Delta t)|\sim J_k(a\Delta t)$ is smaller than a desired
accuracy cutoff. 
This is facilitated by the fast asymptotic decay of the Bessel functions, 
  $J_k(a\Delta t)
  \sim \frac{1}{\sqrt{2\pi k}} \left( \frac{e a\Delta t}{2k} \right)^k$ for 
  $k\to \infty $. Thus, for large $M$, the Chebyshev expansion can be considered as 
quasi-exact. Besides the high accuracy of the method, the linear scaling of 
computation time with both time step and Hilbert space dimension are 
promising in view of potential applications to more complex systems.
In our cases almost all computation time is spent in spMVMs, 
which can be efficiently parallelized, allowing for a good speedup on 
highly parallel computers. \GHcomm{This also means that any
  significant speedup that can be achieved for the MVM, such
  as by our matrix-free formulation, will
  have a corresponding effect on the runtime of the overall
  algorithm. The actual speedup is a function of the memory traffic
  reduction; for instance, a sparse matrix stored in CRS format
  that describes a stencil-like
  neighborhood relation with eight neighbors will (in double precision)
  cause a minimum data traffic of approximately 7.6\,\BF\ when acting on a
  RHS vector. In a matrix-free formulation this balance
  reduces to 1.3\,\BF, leading to a performance improvement of 5.7$\times$
  if the memory bandwidth can be saturated in both cases.}

As an example, we apply the Chebyshev time evolution scheme to the propagation and scattering of a Dirac electron wave packet on a graphene sheet with 
an imprinted gate-defined quantum dot array [\cite{pieper_0295-5075-104-4-47010,Fehske_PSSB:PSSB201552119}]. This is a timely issue of of high experimental relevance [\cite{srep_klein,nc_mie,srep_focusarray}]. We mimic the quantum dot array by implementing the potential $V_n$ in~\eqref{graphene_model}
as
\begin{equation}
V({\bf r}) = \sum_{l=1, k=1}^{L, K} V_k \Theta ( R_{\text {dot}} - |{\bf R}_{l,k} - {\bf r}| )
\label{qdotarray}
\end{equation}
with varying amplitude $V_k = V_0 + \Delta V | k - K/2 |$ in $y$ direction. In~\eqref{qdotarray}, $R_{\text {dot}}$ ($D_\text {dot}$) is the radius of a single quantum dot (the nearest-neighbor distance between dots) and ${\bf R_{l,k}} = ( x_0 + l D_\text {dot}, y_0 + k D_\text {dot} )$  gives the dot's position [$l$ $(k)$ count in $x$ ($y$) direction]. The quantum dot lattice can be created by applying spatially confined top gate voltages. The gradient change of the dot potentials mimics spatially varying effective refraction indices for the Dirac electron waves.

\begin{figure}[tbp]
  \centering\includegraphics[width=0.7\linewidth,clip]{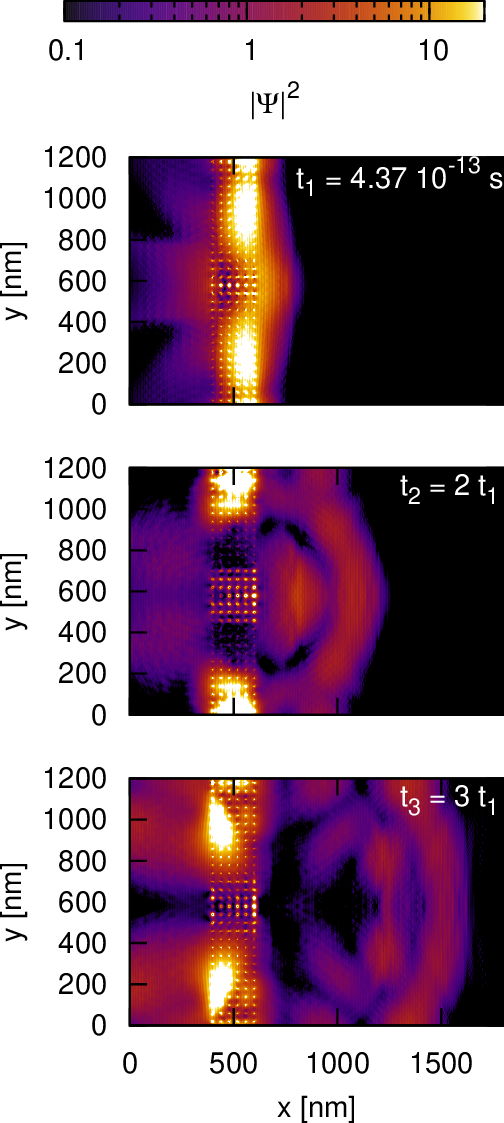}
    \caption{Time evolution of a Dirac electron wave impinging on a graphene quantum dot array (visible by the bright spots).  We consider a GNR (periodic BCs \GHcomm{in $y$ direction}) with an imprinted quantum dot lattice ($L=6$). The radii of the quantum dots $R_{\text {dot}}=10$\,nm, their (midpoint) distance $D_\text {dot}=40$\,nm, and the dot potentials (parameterized by $V_0 = 0.1$, $\Delta V = 0.002$) vary along the $y$ direction between a minimum and maximum value of $y=600$\,nm and $y=0$\,nm or $1200$\,nm, respectively.  
      A (Gaussian) wave packet with momentum in $x$ direction was created at
      \GHcomm{$(x,y)=(200\,\mathrm{nm},600\,\mathrm{nm})$ with 
        $(x,\Delta x)=(200\,\mathrm{nm},300\,\mathrm{nm})$ at time $t=0$.}
      %$(x,y)=(200\,\mathrm{nm},600\,\mathrm{nm})$ at time $t=0$.
      The panels give the (color coded) squared amplitude of the wave function  $|\psi({\bf r},t)|^2$ at times $t_1$, $2t_1$, and $3t_1$ with $t_1=4.37\times 10^{-13}$\,sec (from top to bottom).}
  \label{fig:cheb_tp}
\end{figure}
Figure~\ref{fig:cheb_tp} illustrates the scattering and temporary particle confinement by the quantum dot array. It has been demonstrated by by \cite{Heinisch_RC} and \cite{pieper_0295-5075-104-4-47010} that the normal modes of an isolated quantum dot lead to sharp resonances in the scattering efficiency. Appearing for particular values of $R_{\text {dot}}$, $V$, and $E$, they allow the ``trapping'' even of Dirac electrons. Of course, for the scattering setup considered here, only quasi-bound states can appear, which may have an exceptionally long lifetime, however. Thereby the energy of the wave is fed into vortices inside the dot. For a periodic array of dots the normal modes at neighboring dots can couple, leading to coherence effects (such inter-dot particle transfer takes place on a reduced energy scale compared to pure graphene [\cite{Fehske_PSSB:PSSB201552119}]). The situation becomes more complicated -- but  also more interesting -- when the dot potentials are modulated spatially or energetically.  In this case, a direction-dependent transmission (cascaded Mie scattering [\cite{nc_mie}]) or even the focusing of the electron beam outside the dots can be observed [\cite{srep_focusarray}].  Similar phenomena are demonstrated by Fig.~\ref{fig:cheb_tp}.
\GHcomm{For this simulation the electron is created by a Gaussian wave packet
\begin{equation}
\psi({\bf r},t=0) =\exp\left({-\frac {(x-x_0)^2}{4 \Delta x^2}}\right) \psi_{{\bf K},{\bf x}}({\bf r})~~,
\label{qgauwp}
\end{equation}
where $\psi_{{\bf K},{\bf x}}({\bf r})$ is Dirac electron with momentum in $x$ direction.
}
When the wavefront hits the dot region, the wave is partly trapped by the quantum dots, whereby -- for the parameters used -- the resonance conditions are better fulfilled near the lower/upper end of the dot array (here the particle wave is best captured). The other way around, the transmission (and also the reflection, i.e., the backscattering) is strongest  in the central region, leading to a curved wavefront. For larger time values a second pulse (wavefront)
emerges (note that we have reflections and repeated scattering events in our finite GNR, but the largest time considered, $t=3t_1$, is much shorter than  the pass-through time of an unperturbed Dirac wave). In any case, one observes a strongly time- and direction-dependent emission pattern for the considered graphene-based nanostructure, which can  be exploited to manipulate electron beams. Particularly interesting in this respect would be focusing of the electron beam with large focal length, such that the focal spot lies outside the array. Then the structure can be used as a coupler to other electronic units.  Achieving this by tuning the gradient of the gate potential appears to be a very efficient way, which  is more easily realized in practice than modifying the geometrical properties of the array such as the lattice gradient or the layer number [\cite{srep_focusarray}].

\subsection{Interior eigenvalues of topological insulators}
Since the electronic properties of TIs are mainly determined by the (topologically nontrival) surface states located in the bulk-state gap, an efficient  calculation of electron states at or close to the center of a spectrum is of  vital importance. This can be done by Chebyshev filter diagonalization (ChebFD), a straightforward scheme for  interior eigenvalue computation, which is based on polynomial filter functions and therefore has much in common with the KPM.
ChebFD applies a matrix polynomial filter that is suitable for the
target interval to a block of vectors. In each iteration, the search
space is checked for convergence using a Rayleigh-Ritz procedure.
%The algorithm is thus similar to the power iteration for computing
%extremal eigenvectors.
ChebFD has already proven its practical suitability: Parallelized and implemented on the ``SuperMUC'' supercomputer at LRZ Garching, $10^2$ central eigenvalues of a $10^9$-dimensional sparse matrix have been calculated at 40~Tflop/s sustained performance by \cite{Pieper_ChebFD}.

\begin{figure}%[htbp]
  \centering\includegraphics[width=0.8\linewidth,clip]{dos_topi_disorder}
  \centering\includegraphics[width=0.8\linewidth,clip]{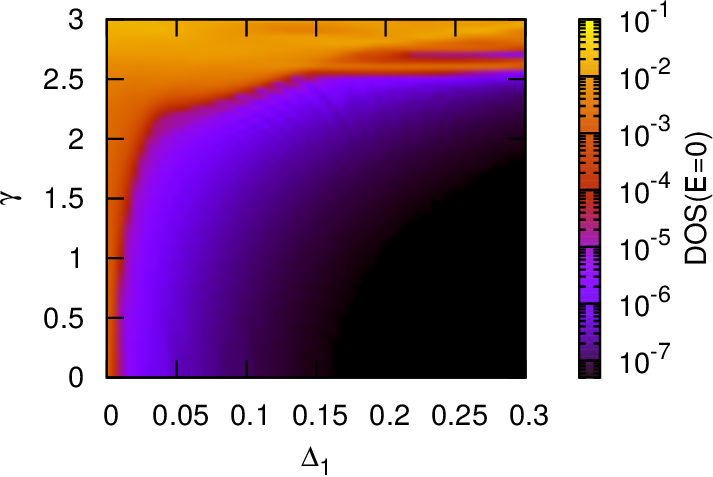}
  \caption{Density of states for a strong TI described by the
    Hamiltonian~\eqref{eq:TB_H_topi} with $m=2$, $\Delta_2=0$, and open
    (periodic) BCs in $z$ ($x$ and $y$) direction.  Top panel: DOS
    without ($V_n=0$; black curve) and with [$V_n\in
    [-\gamma/2,\gamma/2]$; red curve] disorder, where
    $\Delta_1=0$. Data obtained by KPM with stochastic trace
    evaluation for a cuboid with $256 \times 256 \times 10$
    sites. Middle panel: Zoom-in of the central part of the spectrum
    with the target interval used for the ChebFD
    calculations [\cite{Pieper_ChebFD}]. Bottom panel: DOS at the band
    center ($E=0$) in dependence on the gap parameter $\Delta_1$ and
    the disorder strength
    $\gamma$ [\cite{pf2016_Tpoi_PhysRevB.93.035123}]. Applying the KPM,
    2048 Chebyshev moments were used for a system with $512 \times 512
    \times 10$ sites. Note that the finite DOS at $\Delta_1=0$ is a
    finite-size effect and due to the finite KPM resolution (variance
    $\sigma=0.01$).}
  \label{fig:doa_topi}
\end{figure}

Figure~\ref{fig:doa_topi} shows the DOS of a strong TI. The focus is on the (pseudo-) gap region of the DOS. Implementing the effect of nonmagnetic impurities by uniformly distributed random on-site potentials $V_n$, we see how disorder fills the gap that exists in the DOS of system with a finite number of sites (see the red curves in the upper panels). Introducing a  finite $\Delta_1$, which mimics, e.g., the effect of an external magnetic field, the midgap Dirac cone formed by the surface states is broken up. Again, disorder will induce electronic states in the gap region generated by $\Delta_1$. This is demonstrated by the lower panel of Fig.~\ref{fig:doa_topi}, showing the DOS at the band center ($E=0$) in the $\Delta_1$-$\gamma$ plane. As the disorder strength increases, more and more states pop up at $E=0$ until the DOS saturates when $\gamma$ reaches the order of magnitude of the bulk band gap. \GHcomm{For a more detailed investigation of disordered (weak and strong) TI we refer the reader to~\cite{Kobayashi2013} where, besides the phase diagram, also  the DOS was calculated using the KPM (see supplementary material in that paper).}  Compared to KPM, our ChebFD  approach yields a better resolution at the same computational cost in the target interval (band center), which is important regarding the scientific applications.
\begin{table}[tb]
\centering
\begin{tabular}{ r | c c  }
  & test case 1 & test case 2 \\\hline
  size & 480$\times$480$\times$6 &   240$\times$240$\times$6 \\
  eigenpairs & 72 & 40 \\
  Emmy nodes & 8 & 8 \\\hline
  ChebFD: & & \\
   runtime [s] & 2852 & 642 \\
   max res. & $7.3\times 10^{-12}$ & $4.9\times 10^{-15}$ \\\hline
  trLanczosFD: & & \\
   runtime [s] & 760 & 142 \\
   max res. & $3.5\times 10^{-15}$ & $1.7\times 10^{-11}$ \\
\end{tabular}
\caption{\label{tab:fd_cases} Test cases for the filter diagonalization method, using
  a matrix for TI with all eigenvalues in the range [-5.5:5.5]. Runtime and residuum
  data for runs with eight nodes on the Emmy cluster are shown
  for the \final{ChebFD} and the trLanczosFD algorithm, respectively.}
\end{table}

The ChebFD algorithm is robust and scalable, but algorithmically
sub-optimal.  In \pvsctm\ we have also implemented the trLanczosFD
(thick-restart Lanczos with polynomial filters) algorithm 
by \cite{Saad_TRLAwPF}. This algorithm benefits \gh{a little more
  from a matrix-free formulation \final{because it uses smaller block vectors:
  Smaller blocks increase the impact of the data transfer for the matrix
  elements as shown, e.g., in \cite{Kreutzer:2018}. However,
  its actual advantage is improved convergence}}.
A thorough description
would exceed the scope of this paper; in
Table \ref{tab:fd_cases} 
we show runtime data and the maximum residual of the inner Ritz
eigenvalues for trLanczosFD on two TI test cases in comparison with
ChebFD.  TrLanczosFD outperforms ChebFD by a factor of almost four
(using \pvsctm\ for both).

\subsection{Disorder effects in Weyl semimetals}
  
The Weyl nodes in the gapless topological Weyl semimetals are believed to be robust against perturbations unless, e.g., the translation or charge conservation symmetry is broken. Showing the stability of a single or a pair of Weyl nodal points  against disorder has been the subject of intense research [\cite{Liu2016,Chen_weyl_disorder,Pixley_dirac,Cormick_weyl,Hughes_disorder_PhysRevB.93.075108,Zhao_disorder_PhysRevLett.114.206602}]. Due to the vanishing DOS at the nodal points, disorder effects can be expected to be particularly pronounced. Since analytic methods fail widely in their quantitative predictions, even in the case of weak disorder, we use a purely numerical, KPM-based approach to analyze the spectral properties of Weyl semimetals with real-space quenched potential disorder.

\begin{figure}%[htbp]
  \centering\includegraphics[width=0.9\linewidth,clip]{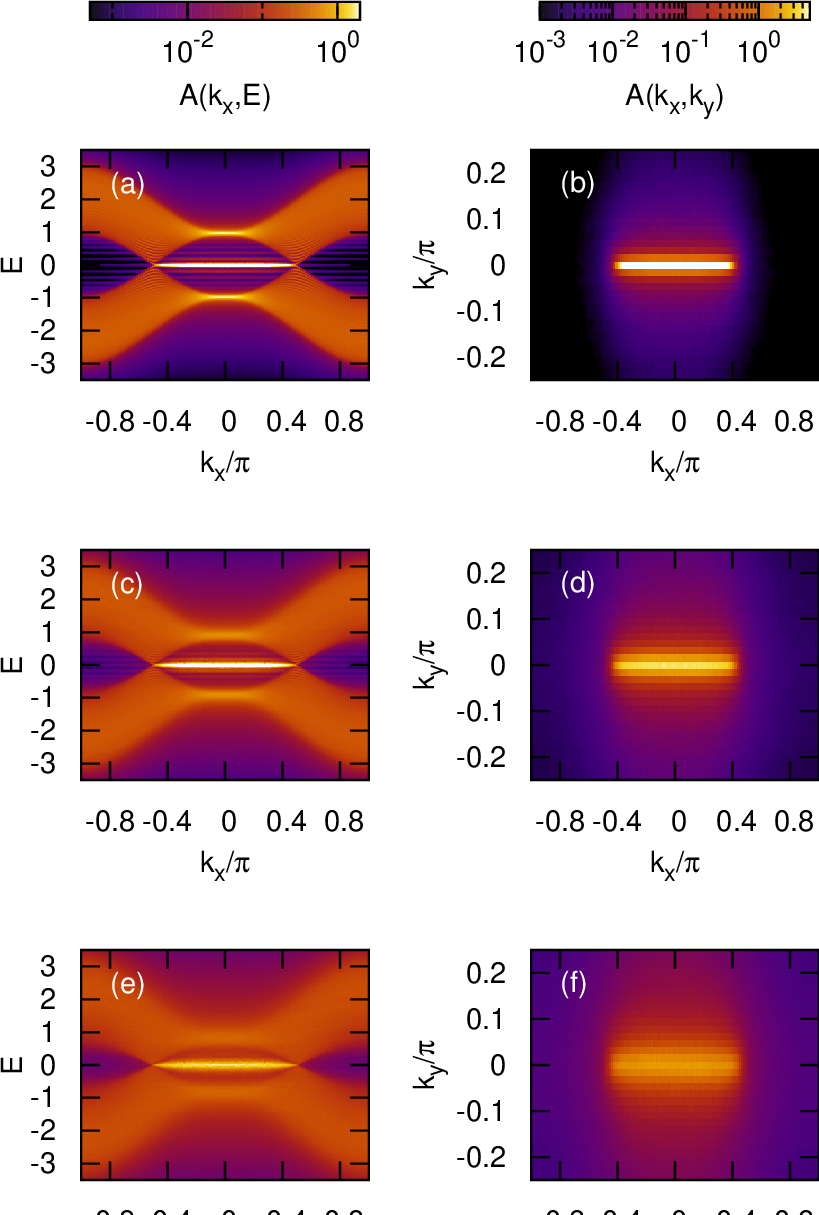}
    \caption{Spectral function $A({\bf k},E)$ for a disordered Weyl semimetal with Fermi arcs, as obtained from the respective equation~\eqref{eq:ak} for a 3D system with $256\times 32\times 32$ sites and periodic (open) BCs in $x,y$ ($z$) directions. The test wave function is initialized only on one surface. Left:  $A({\bf k},E)$  along the $k_x$ direction for $k_y=k_z=0$. Right: $A({\bf k},E)$  in the $k_x$-$k_y$ plane ($k_z=0$) at $E=0$. The disorder strength $\gamma =1$ (top row), $\gamma =2$ (middle row), and $\gamma =3$ (bottom row).}
  \label{fig:msn_aprpes_disorder}
\end{figure}

Figure~\ref{fig:msn_aprpes_disorder} displays the momentum-resolved spectral function $A({\bf k},E)$ of a disordered Weyl metal along different paths in the bulk Brillouin zone. The photoemission spectra shown were calculated for the model~\eqref{eq:TB_H_wsm} with random potentials $V_n$ drawn from  a uniform box distribution of strength $\gamma$, i.e., $V_n \in [-\gamma/2,\gamma/2]$. The presented data  should be compared with the results for the clean case provided by Fig.~\ref{fig:dos_TM_clean}~(c).   Most notably, the electronic states at the Fermi arc (connecting the nodal points) and its immediate vicinity   are hardly influenced by weak and even intermediate disorder. This does not apply for states further away from the Fermi surface. Here, the spectral signatures (band dispersion) are rapidly washed out, even for weak disorder. 
Of course, strong disorder will also affect the Fermi arc and the nodal points: Above a certain disorder strength they will  be smeared out in both energy and momentum space and, as a result, the Weyl semimetal will transform into a diffusive metal with a finite DOS at the nodal points. 
A more detailed investigation of the spectral properties would be desirable in order to confirm the very recent evidence found by \cite{SWW17} for an intermediate Chern insulator state between the disordered Weyl semimetallic and diffusive metallic phases.
At even stronger disorder, the distribution of the local density of states significantly broadens (just as in the case of strongly disordered strong TIs [\cite{SFFV12}] or disordered GNR [\cite{SSF09,schubi_graphene}]) and Anderson localization  sets in [\cite{Pixley_dirac}].

\section{Conclusion and outlook}\label{sec:conc}

The \pvsctm\ DSL and library have been demonstrated to be powerful tools
for generating
high-performance code to investigate ground-state, spectral, and dynamic
properties of graphene, topological insulators, and other materials whose physics is
governed by short-range interactions that lead to stencil-like
numerical kernels.
\gh{In particular, by calculating the time evolution and scattering of
wave packets in graphene-based nanostructures, determining the
interior eigenvalues related to protected surface states in
topological insulators, and treating the effects of random impurities
in Weyl semimetals, we exemplarily showed that the proposed PVSC-DTM
scheme can easily be combined with other numerical algorithms based on
an efficient matrix-vector multiplication and used for studying very
diverse aspects in the highly topical field of functional quantum
matter}.

Due to is matrix-free design, \pvsctm\ outperforms
matrix-based libraries such as GHOST. It also implements effective
SIMD vectorization and fast on-the-fly random number generation and
yields optimal memory-bound chip-level performance as shown by the
roof{}line model. Spatial blocking of the iteration loops is fully
automatic and based on layer conditions.

Several improvements to the DSL library are left for future work: A better
integration of the random number generator with the inner update loop
would increase the non-saturated and sequential performance.
Overlapping computation with communication would improve the
distributed-memory parallel efficiency. Both optimizations are
prerequisites for a possible integration with \gh{advanced blocking
  algorithms. Blocked ChebFD, which we actually used in this work,
  is one example, but more complex schemes exist}.  \gh{Exploiting the
  symmetry of the stencil shape or coefficients is not currently implemented
  in the DSL but could be useful to make writing the code
easier}. The system geometry is currently
limited to rectangular and cuboid domains, \gh{which is
  a restriction \final{that may be lifted
    to support other physical setups, e.g., ring-shaped structures to
    study the Aharonov-Bohm effect or boundary-related (topological) states.}}
  %In the applications considered here,
  %the overall system size
  %is usually chosen much larger than interesting feature sizes
  %such as unit cells.}}
%In that respect, irregular domain shapes
%  are not of foremost interest for the applications addressed by
%  PVSC-DTM.}

Finally we plan to
implement more algorithms in order to make the library more versatile
beyond the showcases described here.

\paragraph*{\bf Acknowledgments} 
We thank  Andreas Al\-ver\-mann, Rafael L. Heinisch, and Gerhard Wellein
for fruitful discussions.
We are grateful for financial support by KONWIHR, the Bavarian Competence
Network for High Performance Computing (projects PVSC-TM and PEigFex). 
We are also indebted to the LRZ Garching for providing access to
HLRB-II\@.

\raggedright


\begin{thebibliography}{61}
\providecommand{\natexlab}[1]{#1}
\providecommand{\url}[1]{\texttt{#1}}
\providecommand{\urlprefix}{URL }
\expandafter\ifx\csname urlstyle\endcsname\relax
  \providecommand{\doi}[1]{DOI:\discretionary{}{}{}#1}\else
  \providecommand{\doi}{DOI:\discretionary{}{}{}\begingroup
  \urlstyle{rm}\Url}\fi

\bibitem[{Alvermann and Fehske(2008)}]{Alver_PhysRevB.77.045125}
Alvermann A and Fehske H (2008) Chebyshev approach to quantum systems coupled
  to a bath.
\newblock \emph{Phys. Rev. B} 77: 045125.
\newblock \doi{10.1103/PhysRevB.77.045125}.

\bibitem[{Alvermann and Fehske(2011)}]{cfet}
Alvermann A and Fehske H (2011) High-order commutator-free exponential
  time-propagation of driven quantum systems.
\newblock \emph{J. Comput. Phys.} 230(15): 5930--5956.
\newblock \doi{10.1016/j.jcp.2011.04.006}.

\bibitem[{Bandishti et~al.(2012)Bandishti, Pananilath and
  Bondhugula}]{Bandishti6468470}
Bandishti V, Pananilath I and Bondhugula U (2012) Tiling stencil computations
  to maximize parallelism.
\newblock In: \emph{Proceedings of the International Conference for High
  Performance Computing, Networking, Storage and Analysis}. pp. 1--11.
\newblock \doi{10.1109/SC.2012.107}.

\bibitem[{{Basu} et~al.(2013){Basu}, {Venkat}, {Hall}, {Williams}, {Van
  Straalen} and {Oliker}}]{Basu:2013}
{Basu} P, {Venkat} A, {Hall} M, {Williams} S, {Van Straalen} B and {Oliker} L
  (2013) Compiler generation and autotuning of communication-avoiding operators
  for geometric multigrid.
\newblock In: \emph{20th Annual International Conference on High Performance
  Computing}. pp. 452--461.
\newblock \doi{10.1109/HiPC.2013.6799131}.

\bibitem[{Caridad et~al.(2016)Caridad, Connaughton, Ott, Weber and
  Krsti\'{c}}]{nc_mie}
Caridad JM, Connaughton S, Ott C, Weber HB and Krsti\'{c} V (2016) An
  electrical analogy to {Mie} scatterings.
\newblock \emph{Nature Communications} 7: 12894.
\newblock \doi{10.1038/ncomms12894}.

\bibitem[{Castro~Neto et~al.(2009)Castro~Neto, Guinea, Peres, Novoselov and
  Geim}]{Graphene_review_geim}
Castro~Neto AH, Guinea F, Peres NMR, Novoselov KS and Geim AK (2009) The
  electronic properties of graphene.
\newblock \emph{Rev. Mod. Phys.} 81: 109--162.
\newblock \doi{10.1103/RevModPhys.81.109}.

\bibitem[{Chamon et~al.(2014)Chamon, Goerbig, Moessner and
  Cugliandolo}]{CGMC17}
Chamon C, Goerbig MO, Moessner R and Cugliandolo LF (eds.)  (2014)
  \emph{Topological Aspects of Condensed Matter Physics}, \emph{Lecture Notes
  of the Les Houches Summer School}, volume 103.
\newblock Oxford University Press.

\bibitem[{Chen et~al.(2015)Chen, Song, Jiang, Sun, Wang and
  Xie}]{Chen_weyl_disorder}
Chen CZ, Song J, Jiang H, Sun Qf, Wang Z and Xie XC (2015) Disorder and
  metal-insulator transitions in {Weyl} semimetals.
\newblock \emph{Phys. Rev. Lett.} 115: 246603.
\newblock \doi{10.1103/PhysRevLett.115.246603}.

\bibitem[{Datta et~al.(2009)Datta, Kamil, Williams, Oliker, Shalf and
  Yelick}]{Datta:2009}
Datta K, Kamil S, Williams S, Oliker L, Shalf J and Yelick K (2009)
  Optimization and performance modeling of stencil computations on modern
  microprocessors.
\newblock \emph{SIAM Review} 51(1): 129--159.
\newblock \doi{10.1137/070693199}.

\bibitem[{Datta et~al.(2008)Datta, Murphy, Volkov, Williams, Carter, Oliker,
  Patterson, Shalf and Yelick}]{Datta:2008}
Datta K, Murphy M, Volkov V, Williams S, Carter J, Oliker L, Patterson D, Shalf
  J and Yelick K (2008) Stencil computation optimization and auto-tuning on
  state-of-the-art multicore architectures.
\newblock In: \emph{Proceedings of the 2008 ACM/IEEE Conference on
  Supercomputing}, SC '08. Piscataway, NJ, USA: IEEE Press.
\newblock ISBN 978-1-4244-2835-9, pp. 4:1--4:12.
\newblock \urlprefix\url{http://dl.acm.org/citation.cfm?id=1413370.1413375}.

\bibitem[{Fehske et~al.(2015)Fehske, Hager and
  Pieper}]{Fehske_PSSB:PSSB201552119}
Fehske H, Hager G and Pieper A (2015) Electron confinement in graphene with
  gate-defined quantum dots.
\newblock \emph{physica status solidi (b)} 252(8): 1868--1871.
\newblock \doi{10.1002/pssb.201552119}.

\bibitem[{Fehske et~al.(2009)Fehske, Schleede, Schubert, Wellein, Filinov and
  Bishop}]{FEHSKE20092182}
Fehske H, Schleede J, Schubert G, Wellein G, Filinov VS and Bishop AR (2009)
  Numerical approaches to time evolution of complex quantum systems.
\newblock \emph{Physics Letters A} 373: 2182 -- 2188.
\newblock \doi{10.1016/j.physleta.2009.04.022}.

\bibitem[{Fu et~al.(2007)Fu, Kane and Mele}]{FKM07}
Fu L, Kane CL and Mele EJ (2007) Topological insulators in three dimensions.
\newblock \emph{Phys. Rev. Lett.} 98: 106803.
\newblock \doi{10.1103/PhysRevLett.98.106803}.

\bibitem[{Goerbig(2011)}]{Go11}
Goerbig MO (2011) Electronic properties of graphene in a strong magnetic field.
\newblock \emph{Rev. Mod. Phys.} 83: 1193--1243.
\newblock \doi{10.1103/RevModPhys.83.1193}.

\bibitem[{Gropp et~al.(2000)Gropp, Kaushik, Keyes and Smith}]{Gropp:1999}
Gropp WD, Kaushik DK, Keyes DE and Smith BF (2000) Towards realistic
  performance bounds for implicit {CFD} codes.
\newblock In: Keyes D, Periaux J, Ecer A, Satofuka N and Fox P (eds.)
  \emph{Parallel Computational Fluid Dynamics 1999}. Elsevier.
\newblock ISBN 978-0-444-82851-4, pp. 241--248.
\newblock \doi{10.1016/B978-044482851-4.50030-X}.

\bibitem[{Hasan and Kane(2010)}]{topi_review_hasan_RevModPhys.82.3045}
Hasan MZ and Kane CL (2010) Colloquium: Topological insulators.
\newblock \emph{Rev. Mod. Phys.} 82: 3045--3067.
\newblock \doi{10.1103/RevModPhys.82.3045}.

\bibitem[{Hasan et~al.(2017)Hasan, Xu, Belopolski and Huang}]{Hasan:2017}
Hasan MZ, Xu SY, Belopolski I and Huang SM (2017) Discovery of {Weyl} fermion
  semimetals and topological {Fermi} arc states.
\newblock \emph{Annual Review of Condensed Matter Physics} 8(1): 289--309.
\newblock \doi{10.1146/annurev-conmatphys-031016-025225}.

\bibitem[{Heinisch et~al.(2013)Heinisch, Bronold and Fehske}]{Heinisch_RC}
Heinisch RL, Bronold FX and Fehske H (2013) Mie scattering analog in graphene:
  Lensing, particle confinement, and depletion of {Klein} tunneling.
\newblock \emph{Phys. Rev. B} 87: 155409.
\newblock \doi{10.1103/PhysRevB.87.155409}.

\bibitem[{Ilic et~al.(2014)Ilic, Pratas and Sousa}]{Loft}
Ilic A, Pratas F and Sousa L (2014) Cache-aware roof{}line model: Upgrading the
  loft.
\newblock \emph{IEEE Computer Architecture Letters} 13(1): 21--24.
\newblock \doi{10.1109/L-CA.2013.6}.

\bibitem[{{Kamil} et~al.(2010){Kamil}, {Chan}, {Oliker}, {Shalf} and
  {Williams}}]{Kamil:2010}
{Kamil} S, {Chan} C, {Oliker} L, {Shalf} J and {Williams} S (2010) An
  auto-tuning framework for parallel multicore stencil computations.
\newblock In: \emph{2010 IEEE International Symposium on Parallel Distributed
  Processing (IPDPS)}. pp. 1--12.
\newblock \doi{10.1109/IPDPS.2010.5470421}.

\bibitem[{Kobayashi et~al.(2013)Kobayashi, Ohtsuki and Imura}]{Kobayashi2013}
Kobayashi K, Ohtsuki T and Imura KI (2013) Disordered weak and strong
  topological insulators.
\newblock \emph{Phys. Rev. Lett.} 110: 236803.
\newblock \doi{10.1103/PhysRevLett.110.236803}.

\bibitem[{Kreutzer et~al.(2018)Kreutzer, Ernst, Bishop, Fehske, Hager, Nakajima
  and Wellein}]{Kreutzer:2018}
Kreutzer M, Ernst D, Bishop AR, Fehske H, Hager G, Nakajima K and Wellein G
  (2018) Chebyshev filter diagonalization on modern manycore processors and
  {GPGPUs}.
\newblock In: Yokota R, Weiland M, Keyes D and Trinitis C (eds.) \emph{High
  Performance Computing}. Cham: Springer International Publishing.
\newblock ISBN 978-3-319-92040-5, pp. 329--349.
\newblock \doi{10.1007/978-3-319-92040-5_17}.

\bibitem[{{Kreutzer} et~al.(2015){Kreutzer}, {Pieper}, {Hager}, {Wellein},
  {Alvermann} and {Fehske}}]{Kreutzer:2015}
{Kreutzer} M, {Pieper} A, {Hager} G, {Wellein} G, {Alvermann} A and {Fehske} H
  (2015) Performance engineering of the {K}ernel {P}olynomal {M}ethod on
  large-scale {CPU}-{GPU} systems.
\newblock In: \emph{2015 IEEE International Parallel and Distributed Processing
  Symposium}. pp. 417--426.
\newblock \doi{10.1109/IPDPS.2015.76}.

\bibitem[{Kreutzer et~al.(2017)Kreutzer, Thies, R{\"o}hrig-Z{\"o}llner, Pieper,
  Shahzad, Galgon, Basermann, Fehske, Hager and Wellein}]{GHOST2016}
Kreutzer M, Thies J, R{\"o}hrig-Z{\"o}llner M, Pieper A, Shahzad F, Galgon M,
  Basermann A, Fehske H, Hager G and Wellein G (2017) {GHOST}: Building blocks
  for high performance sparse linear algebra on heterogeneous systems.
\newblock \emph{International Journal of Parallel Programming} 45(5):
  1046--1072.
\newblock \doi{10.1007/s10766-016-0464-z}.

\bibitem[{L'Ecuyer and Simard(2007)}]{TestU01}
L'Ecuyer P and Simard R (2007) {TestU01}: {A} {C} library for empirical testing
  of random number generators.
\newblock \emph{ACM Trans. Math. Softw.} 33(4): 22:1--22:40.
\newblock \doi{10.1145/1268776.1268777}.

\bibitem[{Li et~al.(2016)Li, Xi, Vecharynski, Yang and Saad}]{Saad_TRLAwPF}
Li R, Xi Y, Vecharynski E, Yang C and Saad Y (2016) A thick-restart {Lanczos}
  algorithm with polynomial filtering for {Hermitian} eigenvalue problems.
\newblock \emph{SIAM Journal on Scientific Computing} 38(4): A2512--A2534.
\newblock \doi{10.1137/15M1054493}.

\bibitem[{Liu et~al.(2016)Liu, Ohtsuki and Shindou}]{Liu2016}
Liu S, Ohtsuki T and Shindou R (2016) Effect of disorder in a three-dimensional
  layered {C}hern insulator.
\newblock \emph{Phys. Rev. Lett.} 116: 066401.
\newblock \doi{10.1103/PhysRevLett.116.066401}.

\bibitem[{Malas et~al.(2017)Malas, Hager, Ltaief and Keyes}]{Girih2017}
Malas TM, Hager G, Ltaief H and Keyes DE (2017) Multidimensional intratile
  parallelization for memory-starved stencil computations.
\newblock \emph{ACM Trans. Parallel Comput.} 4(3): 12:1--12:32.
\newblock \doi{10.1145/3155290}.

\bibitem[{Marsaglia(2003)}]{Xorshift_2003}
Marsaglia G (2003) Xorshift {RNGs}.
\newblock \emph{Journal of Statistical Software} 8(1): 1--6.
\newblock \doi{10.18637/jss.v008.i14}.

\bibitem[{McCalpin(1991-2007)}]{stream}
McCalpin JD (1991-2007) {STREAM}: Sustainable memory bandwidth in high
  performance computers.
\newblock Technical report, University of Virginia, Charlottesville, VA.
\newblock \urlprefix\url{http://www.cs.virginia.edu/stream/}.
\newblock A continually updated technical report.

\bibitem[{McCormick et~al.(2017)McCormick, Kimchi and Trivedi}]{Cormick_weyl}
McCormick TM, Kimchi I and Trivedi N (2017) Minimal models for topological
  {Weyl} semimetals.
\newblock \emph{Phys. Rev. B} 95: 075133.
\newblock \doi{10.1103/PhysRevB.95.075133}.

\bibitem[{Pieper and Fehske(2016)}]{pf2016_Tpoi_PhysRevB.93.035123}
Pieper A and Fehske H (2016) Topological insulators in random potentials.
\newblock \emph{Phys. Rev. B} 93: 035123.
\newblock \doi{10.1103/PhysRevB.93.035123}.

\bibitem[{Pieper et~al.(2013)Pieper, Heinisch and
  Fehske}]{pieper_0295-5075-104-4-47010}
Pieper A, Heinisch RL and Fehske H (2013) Electron dynamics in graphene with
  gate-defined quantum dots.
\newblock \emph{EPL (Europhysics Letters)} 104(4): 47010.
\newblock \doi{10.1209/0295-5075/104/47010}.

\bibitem[{Pieper et~al.(2016)Pieper, Kreutzer, Alvermann, Galgon, Fehske,
  Hager, Lang and Wellein}]{Pieper_ChebFD}
Pieper A, Kreutzer M, Alvermann A, Galgon M, Fehske H, Hager G, Lang B and
  Wellein G (2016) High-performance implementation of {C}hebyshev filter
  diagonalization for interior eigenvalue computations.
\newblock \emph{Journal of Computational Physics} 325: 226--243.
\newblock \doi{10.1016/j.jcp.2016.08.027}.

\bibitem[{Pixley et~al.(2015)Pixley, Goswami and Das~Sarma}]{Pixley_dirac}
Pixley JH, Goswami P and Das~Sarma S (2015) Anderson localization and the
  quantum phase diagram of three dimensional disordered {D}irac semimetals.
\newblock \emph{Phys. Rev. Lett.} 115: 076601.
\newblock \doi{10.1103/PhysRevLett.115.076601}.

\bibitem[{Qi and Zhang(2011)}]{topi_review_Qi_RevModPhys.83.1057}
Qi XL and Zhang SC (2011) Topological insulators and superconductors.
\newblock \emph{Rev. Mod. Phys.} 83: 1057--1110.
\newblock \doi{10.1103/RevModPhys.83.1057}.

\bibitem[{Ragan-Kelley et~al.(2013)Ragan-Kelley, Barnes, Adams, Paris, Durand
  and Amarasinghe}]{Halide}
Ragan-Kelley J, Barnes C, Adams A, Paris S, Durand F and Amarasinghe S (2013)
  Halide: A language and compiler for optimizing parallelism, locality, and
  recomputation in image processing pipelines.
\newblock In: \emph{Proceedings of the 34th ACM SIGPLAN Conference on
  Programming Language Design and Implementation}, PLDI '13. New York, NY, USA:
  ACM.
\newblock ISBN 978-1-4503-2014-6, pp. 519--530.
\newblock \doi{10.1145/2491956.2462176}.

\bibitem[{{Randles} et~al.(2013){Randles}, {Kale}, {Hammond}, {Gropp} and
  {Kaxiras}}]{Randles:2013}
{Randles} AP, {Kale} V, {Hammond} J, {Gropp} W and {Kaxiras} E (2013)
  Performance analysis of the lattice {B}oltzmann model beyond
  {N}avier-{S}tokes.
\newblock In: \emph{2013 IEEE 27th International Symposium on Parallel and
  Distributed Processing}. pp. 1063--1074.
\newblock \doi{10.1109/IPDPS.2013.109}.

\bibitem[{Rivera and Tseng(2000)}]{Rivera:2000}
Rivera G and Tseng CW (2000) Tiling optimizations for {3D} scientific
  computations.
\newblock In: \emph{Supercomputing, ACM/IEEE 2000 Conference}. pp. 32--32.
\newblock \doi{10.1109/SC.2000.10015}.

\bibitem[{Schmitt et~al.(2014)Schmitt, Kuckuk, Hannig, Köstler and
  Teich}]{Exaslang}
Schmitt C, Kuckuk S, Hannig F, Köstler H and Teich J (2014) {ExaSlang}: A
  domain-specific language for highly scalable multigrid solvers.
\newblock In: \emph{2014 Fourth International Workshop on Domain-Specific
  Languages and High-Level Frameworks for High Performance Computing}. pp.
  42--51.
\newblock \doi{10.1109/WOLFHPC.2014.11}.

\bibitem[{Schubert and Fehske(2012)}]{schubi_graphene}
Schubert G and Fehske H (2012) Metal-to-insulator transition and electron-hole
  puddle formation in disordered graphene nanoribbons.
\newblock \emph{Phys. Rev. Lett.} 108: 066402.
\newblock \doi{10.1103/PhysRevLett.108.066402}.

\bibitem[{Schubert et~al.(2012)Schubert, Fehske, Fritz and Vojta}]{SFFV12}
Schubert G, Fehske H, Fritz L and Vojta M (2012) Fate of topological-insulator
  surface states under strong disorder.
\newblock \emph{Phys. Rev. B} 85: 201105.
\newblock \doi{10.1103/PhysRevB.85.201105}.

\bibitem[{Schubert et~al.(2009)Schubert, Schleede and Fehske}]{SSF09}
Schubert G, Schleede J and Fehske H (2009) Anderson disorder in graphene
  nanoribbons: A local distribution approach.
\newblock \emph{Phys. Rev. B} 79: 235116.
\newblock \doi{10.1103/PhysRevB.79.235116}.

\bibitem[{Shapourian and Hughes(2016)}]{Hughes_disorder_PhysRevB.93.075108}
Shapourian H and Hughes TL (2016) Phase diagrams of disordered {Weyl}
  semimetals.
\newblock \emph{Phys. Rev. B} 93: 075108.
\newblock \doi{10.1103/PhysRevB.93.075108}.

\bibitem[{Sitte et~al.(2012)Sitte, Rosch, Altman and Fritz}]{SRAF12}
Sitte M, Rosch A, Altman E and Fritz L (2012) Topological insulators in
  magnetic fields: Quantum {H}all effect and edge channels with a nonquantized
  $\ensuremath{\theta}$ term.
\newblock \emph{Phys. Rev. Lett.} 108: 126807.
\newblock \doi{10.1103/PhysRevLett.108.126807}.

\bibitem[{Stengel et~al.(2015)Stengel, Treibig, Hager and Wellein}]{sthw15}
Stengel H, Treibig J, Hager G and Wellein G (2015) Quantifying performance
  bottlenecks of stencil computations using the {E}xecution-{C}ache-{M}emory
  model.
\newblock In: \emph{Proceedings of the 29th ACM International Conference on
  Supercomputing}, ICS '15. New York, NY, USA: ACM.
\newblock \doi{10.1145/2751205.2751240}.

\bibitem[{Su et~al.(2017)Su, Wang and Wang}]{SWW17}
Su Y, Wang XS and Wang SR (2017) A generic phase between disordered {Weyl}
  semimetal and diffusive metal.
\newblock \emph{Scientific Reports} 7: 14382.
\newblock \doi{10.1038/s41598-017-14760-8}.

\bibitem[{Tal-Ezer and Kosloff(1984)}]{Tal-Ezer1984}
Tal-Ezer H and Kosloff R (1984) An accurate and efficient scheme for
  propagating the time dependent {S}chr{\"o}dinger equation.
\newblock \emph{The Journal of Chemical Physics} 81(9): 3967--3971.
\newblock \doi{10.1063/1.448136}.

\bibitem[{Tang et~al.(2016)Tang, Cao, Guo, Zhang, Che, Yannick, Zhang and
  Du}]{srep_focusarray}
Tang Y, Cao X, Guo R, Zhang Y, Che Z, Yannick FT, Zhang W and Du J (2016)
  Flat-lens focusing of electron beams in graphene.
\newblock \emph{Scientific Reports} 6: 33522.
\newblock \doi{10.1038/srep33522}.

\bibitem[{Tang et~al.(2011)Tang, Chowdhury, Kuszmaul, Luk and
  Leiserson}]{Pochoir}
Tang Y, Chowdhury RA, Kuszmaul BC, Luk CK and Leiserson CE (2011) The {Pochoir}
  stencil compiler.
\newblock In: \emph{Proceedings of the Twenty-third Annual ACM Symposium on
  Parallelism in Algorithms and Architectures}, SPAA '11. New York, NY, USA:
  ACM.
\newblock ISBN 978-1-4503-0743-7, pp. 117--128.
\newblock \doi{10.1145/1989493.1989508}.

\bibitem[{Vigna(2017)}]{Xorshift128plus_2014}
Vigna S (2017) Further scramblings of {M}arsaglia's xorshift generators.
\newblock \emph{Journal of Computational and Applied Mathematics} 315:
  175--181.
\newblock \doi{10.1016/j.cam.2016.11.006}.

\bibitem[{Walls and Hadad(2015)}]{srep_klein}
Walls J and Hadad D (2015) Suppressing {Klein} tunneling in graphene using a
  one-dimensional array of localized scatterers.
\newblock \emph{Scientific Reports} 5.
\newblock \doi{10.1038/srep08435}.

\bibitem[{Wan et~al.(2011)Wan, Turner, Vishwanath and Savrasov}]{Wan_weyl}
Wan X, Turner AM, Vishwanath A and Savrasov SY (2011) Topological semimetal and
  {Fermi}-arc surface states in the electronic structure of pyrochlore
  iridates.
\newblock \emph{Phys. Rev. B} 83: 205101.
\newblock \doi{10.1103/PhysRevB.83.205101}.

\bibitem[{Wei{\ss}e and Fehske(2008)}]{WF08a}
Wei{\ss}e A and Fehske H (2008) Exact diagonalization techniques.
\newblock In: Fehske H, Schneider R and Wei{\ss}e A (eds.) \emph{Computational
  Many-Particle Physics}, \emph{Lecture Notes in Physics}, volume 739. p. 529.
\newblock \doi{10.1007/978-3-540-74686-7_18}.

\bibitem[{Wei\ss{}e et~al.(2006)Wei\ss{}e, Wellein, Alvermann and Fehske}]{kpm}
Wei\ss{}e A, Wellein G, Alvermann A and Fehske H (2006) The {K}ernel
  {P}olynomial {M}ethod.
\newblock \emph{Rev. Mod. Phys.} 78: 275--306.
\newblock \doi{10.1103/RevModPhys.78.275}.

\bibitem[{Williams et~al.(2009)Williams, Waterman and
  Patterson}]{roofline:2009}
Williams S, Waterman A and Patterson D (2009) Roof{}line: An insightful visual
  performance model for multicore architectures.
\newblock \emph{Commun. ACM} 52(4): 65--76.
\newblock \doi{10.1145/1498765.1498785}.

\bibitem[{Wonnacott(2000)}]{Wonnacott:2000}
Wonnacott DG (2000) Using time skewing to eliminate idle time due to memory
  bandwidth and network limitations.
\newblock In: \emph{International Parallel and Distributed Processing
  Symposium}. pp. 171--180.
\newblock \doi{10.1109/IPDPS.2000.845979}.

\bibitem[{Xu et~al.(2015)Xu, Belopolski, Alidoust, Neupane, Bian, Zhang,
  Sankar, Chang, Yuan, Lee, Huang, Zheng, Ma, Sanchez, Wang, Bansil, Chou,
  Shibayev, Lin, Jia and Hasan}]{Xu613}
Xu SY, Belopolski I, Alidoust N, Neupane M, Bian G, Zhang C, Sankar R, Chang G,
  Yuan Z, Lee CC, Huang SM, Zheng H, Ma J, Sanchez DS, Wang B, Bansil A, Chou
  F, Shibayev PP, Lin H, Jia S and Hasan MZ (2015) Discovery of a {Weyl}
  fermion semimetal and topological {Fermi} arcs.
\newblock \emph{Science} 349(6248): 613--617.
\newblock \doi{10.1126/science.aaa9297}.

\bibitem[{Yang et~al.(2011)Yang, Lu and Ran}]{Yang_weyl}
Yang KY, Lu YM and Ran Y (2011) Quantum {Hall} effects in a {Weyl} semimetal:
  Possible application in pyrochlore iridates.
\newblock \emph{Phys. Rev. B} 84: 075129.
\newblock \doi{10.1103/PhysRevB.84.075129}.

\bibitem[{Zhang et~al.(2017)Zhang, Driscoll, Markley, Williams, Basu and
  Fox}]{Snowflake}
Zhang N, Driscoll M, Markley C, Williams S, Basu P and Fox A (2017) Snowflake:
  A lightweight portable stencil {DSL}.
\newblock In: \emph{2017 IEEE International Parallel and Distributed Processing
  Symposium Workshops (IPDPSW)}. pp. 795--804.
\newblock \doi{10.1109/IPDPSW.2017.89}.

\bibitem[{Zhao and Wang(2015)}]{Zhao_disorder_PhysRevLett.114.206602}
Zhao YX and Wang ZD (2015) Disordered {Weyl} semimetals and their topological
  family.
\newblock \emph{Phys. Rev. Lett.} 114: 206602.
\newblock \doi{10.1103/PhysRevLett.114.206602}.

\end{thebibliography}
\end{document}